\documentclass[12pt]{article}
\usepackage{graphicx} 
\usepackage{amsmath}
\usepackage[dvipsnames]{xcolor}
\usepackage{color}
\usepackage{braket}
\usepackage{tikz}
\usepackage{physics}
\usepackage{svg}
\usepackage{dsfont}
\usepackage{hyperref}
\usepackage{amsfonts}
\usepackage{comment}
\usepackage{amssymb}

\renewcommand{\thefootnote}{\fnsymbol{footnote}}

\newcommand{\ie}{{\it i.e.}}
\newcommand{\eg}{{\it e.g.}}

\numberwithin{equation}{section}

\begin{document}


\begin{titlepage}
\begin{centering}
{\boldmath\bf \Large Tests of restricted quantum focusing and a new CFT bound}

\vspace{6mm}

{\bf Victor Franken$^1$$^2$\footnote{victor.franken@ugent.be}, Sami Kaya$^3$\footnote{samikaya@berkeley.edu}, Fran\c{c}ois Rondeau$^4$\footnote{francois.rondeau@ens-lyon.fr}, Arvin Shahbazi-Moghaddam$^3$$^5$\footnote{arvinshm@gmail.com} and Patrick Tran$^3$\footnote{patrick.tran@berkeley.edu}}

 \vspace{3mm}

$^1$ {\em CPHT, CNRS, Ecole polytechnique, Institut Polytechnique de Paris\\
91120 Palaiseau, France }

$^2$ {\em Department of Physics and Astronomy, Ghent University,\\
Krijgslaan, 281-S9, 9000 Gent, Belgium}

$^3$ {\em Leinweber Institute for Theoretical Physics and Department of Physics, University of California,
 Berkeley, CA 94720 USA}

$^4$ {\em ENS de Lyon, Laboratoire de Physique, CNRS UMR 5672, Lyon 69007, France}

$^5$ {\em Leinweber Institute for Theoretical Physics, Stanford University, Stanford, CA 94305}

\end{centering}
\vspace{0.5cm}
$~$\\
\centerline{\bf\large Abstract}\vspace{0.2cm}

\vspace{-0.6cm}
\begin{quote}

The restricted quantum focusing conjecture (rQFC) plays a central role in an axiomatic formulation of semiclassical gravity. Since much hinges on its validity, it is imperative to subject the rQFC to rigorous tests in novel settings. Here we do so in two independent directions.

First, we prove rQFC in a class of spacetime dimension $d=2$ toy models, JT gravity coupled to a QFT. We also construct explicit counter-examples to the original and \emph{stronger} Quantum Focusing Conjecture in a regime where matter quantum effects are comparable to the total dilaton value.

Second, for $d>2$, we derive from the rQFC a constraint stronger than the Quantum Null Energy Condition (QNEC). In a broad class of states, this bound forbids the QNEC from saturating faster than $O(\mathcal{A})$ as the transverse area $\mathcal{A}$ of a certain null deformation shrinks to zero. We speculate about a universal strengthened QNEC holding across all QFT states.

\end{quote}

\end{titlepage}
\setcounter{footnote}{0}
\renewcommand{\thefootnote}{\arabic{footnote}}
 \setlength{\baselineskip}{.7cm} \setlength{\parskip}{.2cm}

\newpage
\tableofcontents

\section{Introduction}

Semiclassical gravity is believed to emerge from a fundamental theory of quantum gravity. To learn about this fundamental theory, a powerful strategy is to find imprints of its basic principles in the semiclassical limit. Take, for example, the principle of independence between different fundamental subsystems. In Anti-de Sitter (AdS)/Conformal Field Theory (CFT) correspondence, where fundamental theory is the boundary CFT on some Lorentzian manifold, spacelike separated domains of dependence are independent. The semiclassical imprint of this in AdS is that the gravity duals of the boundary domains of dependence must also be spacelike separated. This gravity constraint was in turn shown to follow from a deeper quantum gravity constraint called the \emph{quantum focusing conjecture} (QFC).

The QFC was first proposed in~\cite{Bousso:2015mna} and has since been crucially used in proofs of many significant results in AdS/CFT, including the above-mentioned ``entanglement wedge nesting'' (see \eg, \cite{Akers:2016ugt,Akers:2017ttv, Brown:2019rox,Engelhardt:2021qjs, May:2019odp,Engelhardt:2024hpe, Bousso:2024iry}). Beyond AdS/CFT, the QFC implies the generalized second law (GSL)~\cite{Bekenstein:1972tm, Bousso:2015mna}, quantum singularity theorems~\cite{Wall:2010jtc, Bousso:2022tdb, Bousso:2025xyc}, and the existence of quantum extremal surfaces~\cite{Engelhardt:2014gca, Akers:2016ugt, Akers:2019lzs}, leading in particular to a resolution of the black hole information problem~\cite{Penington:2019npb,Almheiri:2019psf}. The QFC also implies the Quantum Null Energy Condition (QNEC), a bound in Quantum Field Theory (QFT) which was then proved with various different QFT techniques \cite{Bousso:2015wca,Koeller:2015qmn,Akers:2017ttv, Balakrishnan:2017bjg,Ceyhan:2018zfg, Hollands:2025glm}.

\begin{figure}
    \centering
    \includegraphics[width=0.5\linewidth]{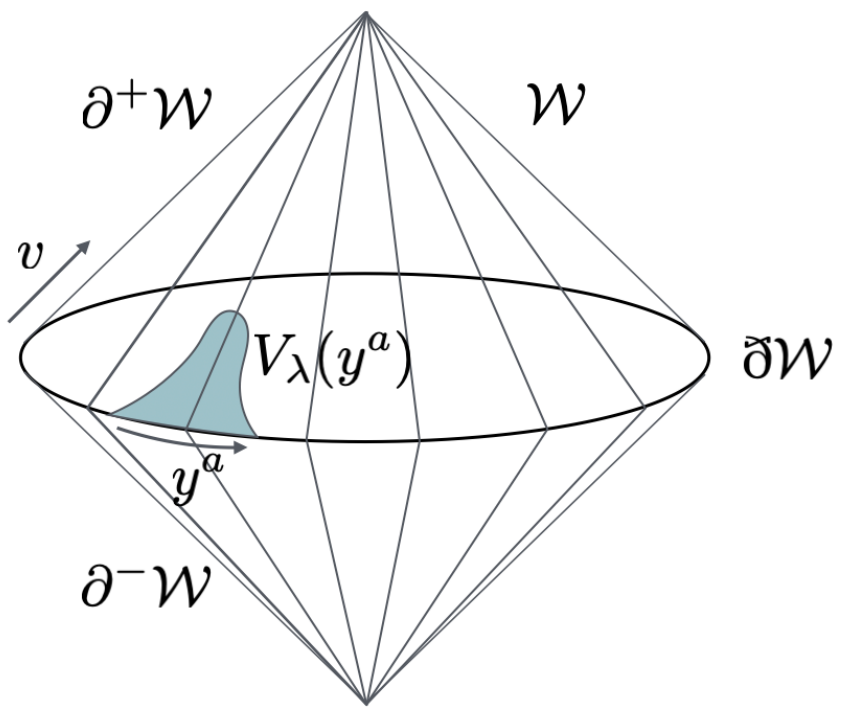}
    \caption{A spacetime wedge $\mathcal{W}$ with future and past boundaries $\partial^\pm \mathcal{W}$ is shown. The (restricted) QFC, roughly, constrains the second derivative of the generalized entropy of spacetime wedges under deformations in which the edge, $\eth \mathcal{W} = \partial^+\mathcal{W} \cap \partial^- \mathcal{W}$, is deformed along the generators of $\partial^+\mathcal{W}$ to a new location $v = V_\lambda(y^a)$, as shown in shaded blue.}
    \label{fig:expansion}
\end{figure}

To state the QFC, let us first introduce its main ingredient, the generalized entropy $S_{\text{gen}}$, a sort of entropy of spacetime regions~\cite{PhysRevD.7.2333}. Let $\mathcal{W}$ denote a spacetime wedge (\ie, a domain of dependence of a partial Cauchy slice) in a general spacetime. In perturbative quantum (Einstein) gravity, the generalized entropy of $\mathcal{W}$ is schematically given by:
\begin{align}\label{eq-13r3r}
    S_{\text{gen}}(\mathcal{W}) = \frac{A(\eth \mathcal{W})}{4G} + S_{\rm ren}(\mathcal{W})+\cdots,
\end{align}
where $A(\eth \mathcal{W})$ denotes the area of the \emph{edge} of $\mathcal{W}$, defined as $\eth \mathcal{W} = \partial^+ \mathcal{W} \cap \partial^- \mathcal{W}$ where $\partial^\pm \mathcal{W}$ denote the future/past sections of $\partial \mathcal{W}$, the boundary of $\mathcal{W}$ (see Fig.~\ref{fig:expansion}), and $S_{\rm ren}(\mathcal{W})$ denotes the renormalized von Neumann entropy of the fields in $\mathcal{W}$. In fact, the complete generalized entropy functional involves more terms than in Eq.~\eqref{eq-13r3r}. Later on, in Sec.~\ref{sect:semi}, we will discuss the complete expansion of $S_{\rm gen}$ in the context of semiclassical gravity.

The QFC constrains the variations of $S_{\text{gen}}$ under certain deformations of $\mathcal{W}$. In particular, let us deform the wedge by translating $\eth \mathcal{W}$ in a neighborhood of a point on $\eth \mathcal{W}$ along the null generators of $\partial^+\mathcal{W}$ (which emanate orthogonally from $\eth \mathcal{W}$) by an affine parameter amount $V(y^a)$, where $y^a$ denotes coordinates for the transverse direction $\eth \mathcal{W}$, so $a\in \{1,\cdots,d-2\}$ where $d$ denotes spacetime dimensionality.

The \emph{quantum expansion} of any wedge in this $V(y^a)$-labeled family is a function $\left.\Theta\right\rvert_\mathcal{W}:\eth \mathcal{W} \to \mathbb{R}$ defined by
\begin{align}\label{eq-11cx5}
    \left.\Theta\right\rvert_\mathcal{W}(y^a) = \frac{4G}{\sqrt{h}}\frac{\delta S_{\text{gen}}}{\delta V(y^a)},
\end{align}
where $h$ denotes the determinant of the intrinsic metric of $\eth \mathcal{W}$~\cite{Bousso:2015mna}.\footnote{We suppose here that $V(y^a)$ is small enough that one does not encounter caustics along $\partial^+ \mathcal{W}$. This can always be arranged for smooth $\eth \mathcal{W}$ which we restrict to for the rest of this paper.}

Let us now define a $1$-parameter family of wedges $\mathcal{W}_\lambda$ by $V_\lambda(y^a)$ such that $\partial_\lambda V_\lambda(y^a) \geq 0$. The QFC is then the following constraint:
\begin{align}\label{eq-3c4134}
    \partial_\lambda \Theta_\lambda \leq 0,
\end{align}
where the notation $\Theta_\lambda$ will be our shorthand for $\left.\Theta\right\rvert_{\mathcal{W}_\lambda}$. Note that the QFC demands Eq.~\eqref{eq-3c4134} on all of $\eth{\mathcal{W}_\lambda}$.

Despite its central role in quantum gravity, the QFC has not been proven. In~\cite{Shahbazi-Moghaddam:2022hbw}, a weaker constraint called the \emph{restricted quantum focusing conjecture} (rQFC) was put forth. The rQFC states:
\begin{align}\label{eq-324c}
      \partial_\lambda \Theta_\lambda(y^a) \leq 0, ~~~~\text{at $\lambda$ and $y^a \in \eth \mathcal{W}_\lambda$ such that $\Theta_\lambda(y^a)= 0$}.
\end{align}
Even though rQFC is weaker than the QFC, it was argued in~\cite{Shahbazi-Moghaddam:2022hbw} that it implies all crucial known implications of the QFC, including the ones mentioned above.\footnote{This is because, so far, in all applications only the following implication of the QFC is used:
\begin{align}\label{eq-2c14}
\Theta_{\lambda_1} \leq 0 \implies \Theta_{\lambda_2}\leq0,\qquad \text{if   }\lambda_1\leq\lambda_2,
\end{align}
where $\Theta_\lambda \leq 0$ means everywhere in the transverse direction $\eth \mathcal{W}_\lambda$. Quite generically, a violation of Eq.~\eqref{eq-2c14} implies the existence of a $\lambda$ between $\lambda_1$ and $\lambda_2$, at which Eq.~\eqref{eq-324c} is violated. Highly non-generic cases (e.g., $\Theta_\lambda = \lambda^3$, where $\partial_\lambda \Theta_\lambda = \partial_\lambda^2 \Theta_\lambda=0$ at $\lambda=0$) naively evades this argument as it satisfies Eq.~\eqref{eq-324c} but not Eq.~\eqref{eq-2c14}. We believe that these non-generic examples are also disallowed since otherwise a slight perturbation to these spacetimes, or even to the family $V_\lambda(y^a)$ will bring back genericity and would violate Eq.~\eqref{eq-324c}. To skip this subtlety entirely, one can \emph{define} rQFC to be the statement of Eq.~\eqref{eq-2c14}, as was done for example in~\cite{Akers:2023fqr, Bousso:2024iry}. We thank Raphael Bousso and Edward Witten for bringing up this non-generic case to our attention.} Furthermore, the rQFC was proved in~\cite{Shahbazi-Moghaddam:2022hbw} in a class of semiclassical gravity theories, the so-called holographic braneworld models~\cite{Randall:1999ee,Randall:1999vf,Shiromizu:1999wj,Gubser:1999vj,Verlinde:1999xm,Karch:2000ct,Emparan:2006ni,Myers:2013lva}.\footnote{The proof strategy in~\cite{Shahbazi-Moghaddam:2022hbw} is not obviously generalizable to a proof of $\partial_\lambda \Theta_\lambda \leq 0$.} The rQFC has since been adopted as a central piece of an axiomatic framework of semiclassical gravity~\cite{Bousso:2024iry, Akers:2023fqr}. We therefore find it necessary to put the rQFC to new tests, to both check its validity and learn new things from it.
Along these lines, we can ask two natural questions:
\begin{itemize}
    \item Does the rQFC hold in all reasonable theories? And, are there violations of the (original) QFC, especially as its crucial known implications appear to simply follow from the weaker rQFC?
    \item Can we predict novel QFT constraints from the rQFC? (A direct proof of which would then constitute non-trivial evidence for the rQFC.)
\end{itemize}
  
This paper is motivated by these questions. In particular, in Sec.~\ref{sect:JT} we prove the rQFC in a class of $d=2$ toy models, Jackiw-Teitelboim (JT) gravity coupled to a QFT. We then find explicit counter-examples to the original QFC in this model. The violations occur in a regime where we do not know of a dimensional reduction interpretation of the toy model, and must therefore be understood strictly within the toy model. Consequently, we do not hold this as very convincing evidence against the QFC. It is remarkable, however, that the rQFC seems to be robust enough to hold in that regime. In Sec.~\ref{sect:CFT}, we focus on $d>2$, and show that a new $G \to 0$ limit of the rQFC, one in which we simultaneosly send the width of the null deformation profile to zero, implies a novel CFT constraint. As we review below, the standard $G \to 0$ limit was used to extract the QNEC. In the class of states we consider, the novel constraints are indeed stronger than the QNEC.

Before summarizing these results in more detail, let us say that the above-mentioned definitions of QFC/rQFC for general $\mathcal{W}$ require the existence of a classical background metric.\footnote{It is conceivable that the requirement of an exactly classical metric can be relaxed, at least for certain region such as exteriors (or interior) of event horizons. And therefore our use of the semiclassical regime is more for caution than necessity. See Sec.~\ref{sect:semi} for a discussion.} In order to obtain a classical background at a level of approximation in which quantum corrections from the matter sector are important in the generalized entropy, a standard approach is to consider a \emph{semiclassical gravity} regime~\footnote{So far in the introduction, as in much of the literature, the term semiclassical gravity was used vaguely. Here, and in Sec.~\ref{sect:semi} we attempt to clarify our use of the term in this paper.} in which large number of matter fields $c$ are present and one takes the limit:
\begin{align}\label{eq-1c24156c}
    G \to 0, ~~ c G \equiv \ell_S^{d-2} = \text{fixed},
\end{align}
where we call the scale $\ell_S$ controlling the backreaction the \emph{species scale}. This regime in effect suppresses quantum corrections from gravitons compared to those from the matter sector, making it possible to approximate the spacetime as classical and backreacting to $\langle T_{ij} \rangle$, the expectation value of the stress-energy tensor.

Given our use of the semiclassical regime, we will review it in more detail in Sec.~\ref{sect:semi}. In the rest of this section, we will summarize the main results of this paper, which are divided into two separate and self-contained sections \ref{sect:JT} and \ref{sect:CFT}.

\subsection*{Results in $d=2$ toy model: rQFC proof and QFC counter-example}

In Sec.~\ref{sect:JT}, we investigate the validity of the QFC and rQFC in JT gravity coupled to a two-dimensional QFT with large number of species in the semiclassical regime~\eqref{eq-1c24156c}.\footnote{Even though this model arises approximately as a dimensional reduction of higher dimensional gravity theories, we treat it as an exact $d=2$ toy model, and do not focus on any higher dimensional interpretation, except for some sporadic remarks.} In this theory, we define the quantum expansion as:
\begin{align}\label{eq-23c1412}
    \Theta_\lambda = \frac{4 G}{\Phi}\partial_\lambda S_\text{gen}.
\end{align}
This definition is informed by the more general definition in Eq.~\eqref{eq-11cx5}, where we have replaced the transverse area factor $\sqrt{h}$ with its JT gravity counterpart, the dilaton $\Phi$.\footnote{See Appendix \ref{sec:reduc} for a review of this dictionary.}

In this model, the rQFC follows from the QNEC, which we will discuss in a moment, and which was proven in QFT in \cite{Bousso:2015wca,Koeller:2015qmn,Balakrishnan:2017bjg,Ceyhan:2018zfg, Hollands:2025glm}, as well as in two-dimensional CFTs in \cite{Almheiri:2019yqk}. Taking the derivative of $\Theta_{\lambda}$ and using the QNEC, we find
\begin{equation}
\label{eq:bd}
    \partial_{\lambda}\Theta_{\lambda} \leq - \theta_{\lambda}\Theta_{\lambda},
\end{equation}
where $\theta_\lambda = \partial_\lambda \log \Phi$. It follows that 
\begin{equation}
    \Theta_{\lambda}=0 \hspace{1em} \Rightarrow \hspace{1em}\partial_{\lambda}\Theta_{\lambda}\leq 0.
\end{equation}
This shows that the rQFC holds while allowing for QFC violations when $\theta_{\lambda}\Theta_{\lambda} < 0$. It might be naively tempting to say that the bound~\eqref{eq:bd} which is stronger than rQFC but not equivalent to the QFC holds in quantum gravity in general dimensions. This is false, however, since the bound in the classical regime amounts to $\partial_\lambda^2 (\text{Area}) \leq 0$ and is therefore violated by a lightcone in Minkowski spacetime whose cross sectional area, say in $d=4$, satisfies $\text{Area} = 4 \pi \lambda^2$.

The central part of our rQFC proof is the use of the two-dimensional QNEC. This proof strategy is identical to the work of Almheiri-Mahajan-Maldacena (AMM) in~\cite{Almheiri:2019yqk}, and in that sense it is not original. There is, however, a crucial distinction between the interpretation of our results. AMM present their work as a proof of the QFC, whereas we derive the rQFC. The discrepancy between the claims stems from the fact that AMM define quantum expansion without the factor of $\Phi$ in the denominator in Eq.~\eqref{eq-23c1412}, a choice which breaks with the original definition of the quantum expansion which is used in defining the QFC. With the proper definition (Eq.~\eqref{eq-23c1412}), their proof simply reduces to ours, as we present in Sec.~\ref{sect:JT}.

In the remainder of Sec.~\ref{sec:QFCviolations}, we study explicit scenarios in our toy model with negative and positive cosmological constant and find explicit violations of the QFC (Eq.~\eqref{eq-11cx5}). We briefly assess the regimes and regions where we should look for such violations. The bound~\eqref{eq:bd} shows that counter-examples can only appear in regions where matter quantum effects are competing with the scale of the dilaton ($\ell_S=cG \gtrsim \Phi$). Intuitively, the sufficient condition $\theta_{\lambda}\Theta_{\lambda}<0$ implies that the variation of matter entropy is strong enough to switch the direction of quantum expansion compared to the direction of geometric expansion. Such regions exist close to evaporating horizons.

Motivated by this observation, we study the near-horizon regions of an AdS extremal black hole and of an evaporating cosmological horizon in de Sitter spacetime. In particular, we construct these backgrounds in JT gravity coupled to a CFT. Using a strong version of the QNEC derived in \cite{Wall:2011kb}, one can show that violations only appear if the species scale $cG$ is of the order or greater than the dilaton $\Phi$, that is, if the quantum effects of matter on the quantum expansion are allowed to compete with the geometric terms. In this regime, we then explicitly find large subregions where $\theta_{\lambda}\Theta_{\lambda}<0$, and for which the QFC is indeed violated.

This is a regime where JT gravity does not admit a higher dimensional uplift, and is therefore in a true sense a toy model. We therefore do not interpret this as strong evidence against the original QFC.

\subsection*{Results in $d>2$: rQFC implies a new CFT bound}

In Sec.~\ref{sect:CFT}, we show that in a class of near-vacuum states, the rQFC implies a new CFT bound. This is a new instance of quantum gravity constraints predicting QFT constraints. A well-known previous example is the QNEC, which was originally derived in~\cite{Bousso:2015mna} from the QFC, and subsequently proved in several papers~\cite{Bousso:2015wca,Koeller:2015qmn,Balakrishnan:2017bjg,Ceyhan:2018zfg, Hollands:2025glm}. The new CFT bound is stronger than the QNEC.


Let us demonstrate how the QNEC is implied by the rQFC.\footnote{As opposed to the QFC, which is an unnecessarily stronger condition for this purpose.} This will set the stage for the above-mentioned new CFT constraint. Here, we will work with the schematic form of the generalized entropy in Eq.~\eqref{eq-13r3r}, namely just the area and the renormalized entropy term. In Sec.~\ref{sect:CFT} we will be more careful and reproduce the main results using a more careful ansatz of the perturbative semiclassical regime (see Sec.~\ref{sect:semi}), including the subleading corrections which will be omitted here. From Eq.~\eqref{eq-13r3r}, we obtain
\begin{align}\label{eq-06932}
    \Theta_\lambda = \theta_\lambda + \frac{4 G}{\sqrt{h_\lambda}} \left.\frac{\delta S_{\rm ren}}{\delta V(y^a)}\right\rvert_\lambda + \cdots,
\end{align}
where $\theta_\lambda = \partial_\lambda \log\sqrt{h_\lambda}$ (where $h_\lambda$ is the determinant of the intrinsic metric of $\eth \mathcal{W}_\lambda$) is the classical expansion of the $V_\lambda$ congruence. The ellipsis here and in the rest of this section is there to indicate we are dropping corrections, keeping only terms that we claim will be relevant for the argument.

We can now take another $\lambda$ derivative of Eq.~\eqref{eq-06932} to obtain:
\begin{align}\label{eq-06933}
    \partial_\lambda \Theta_\lambda &= - \frac{\theta_\lambda^2}{d-2}-\varsigma_\lambda^2- R_{ij} k^i k^j - \frac{4G \theta_\lambda}{\sqrt{h_\lambda}} \left.\frac{\delta S_{\rm ren}}{\delta V(y^a)}\right\rvert_\lambda + \frac{4 G}{\sqrt{h_\lambda}} \partial_\lambda \left.\frac{\delta S_{\rm ren}}{\delta V(y^a)}\right\rvert_\lambda + \cdots,
\end{align}
where $\varsigma_\lambda^2$ is the shear-squared of $\partial^+\mathcal{W}$. The first three terms on the RHS of Eq.~\eqref{eq-06933} equal $\partial_\lambda \theta_\lambda$ through the Raychaudhuri's equation for the affine (surface-orthogonal) congruence $V_\lambda$. The last two terms in Eq.~\eqref{eq-06933} both come from the $\lambda$ derivative of the second term in Eq.~\eqref{eq-06932}.

Now, consider a QFT state in $d>2$ Minkowski spacetime, and let $\mathcal{W}_\lambda$ be a null-deformed family starting from the domain of dependence of a ball-shaped region.\footnote{More precisely, we mean that the wedge $\mathcal{W}$ approaches the domain of dependence of a ball-shaped region in the $G \to 0$ limit. For a general state, the region will get perturbatively distorted away from this.} The spacetime will in general backreact to the state, in particular, through the following component of Einstein field equations:
\begin{align}\label{eq-3214122}
    R_{ij} k^i k^j = 8 \pi G \langle T_{ij}\rangle k^i k^j,
\end{align}
where $\langle T_{ij} \rangle$ is the expectation value of the stress-energy tensor, and $k^i$ is any null vector.\footnote{Coupling the expectation value $\langle T_{ij} \rangle$ to the classical field $R_{ij}$ is justified in the semiclassical gravity regime \eqref{eq-1c24156c}.}

Next, we impose the rQFC condition $\Theta_\lambda(y^a) = 0$ at some point $y^a \in \eth \mathcal{W}$, perturbatively in $G$:
\begin{align}\label{eq-13251}
    \Theta_\lambda(y^a) = 0 \implies \theta_\lambda(y^a) =- \frac{4 G}{\sqrt{h_\lambda}} \left.\frac{\delta S_{\rm ren}}{\delta V(y^a)}\right\rvert_\lambda + \cdots.
\end{align}
This enforces the (un-backreacted) size of the ball $R$ to scale with $G$, because $\theta_\lambda \sim R^{-1}$.

Finally, by combining Eqs.~\eqref{eq-06933}, \eqref{eq-3214122}, and \eqref{eq-13251}, we get:
\begin{equation}\label{eq-24t1}
\begin{split}
    \Theta_\lambda(y^a)=0 \implies \partial_\lambda \Theta_\lambda(y^a) =&\,\frac{16(d-3)G^2}{(d-2)} \left(\left.\frac{1}{\sqrt{h_\lambda}}\frac{\delta S_{\rm ren}}{\delta V(y^a)}\right\rvert_\lambda\right)^2\\
    &- 4 G \left( 2\pi \langle T_{ij}\rangle k^i k^j -\frac{1}{\sqrt{h_\lambda}}\partial_\lambda \left.\frac{\delta S_{\rm ren}}{\delta V(y^a)}\right\rvert_{\lambda}\right) + \cdots,
\end{split}
\end{equation}
where $k^i$ is the (surface-orthogonal) null vector field on $\partial^+\mathcal{W}$ which generates the $V_\lambda(y^a)$ flow. Note that since the shear of the $\partial^+\mathcal{W}$ congruence is zero in Minkowski spacetime, and the perturbation to Minkowski space is $O(G)$, therefore the shear-squared term in Eq.~\eqref{eq-06933} is $O(G^2)$.

Demanding the non-positivity of $\partial_\lambda \Theta_\lambda$ in Eq.~\eqref{eq-24t1} is a re-statement of the gravitational constraint rQFC.\footnote{The expression in Eq.\eqref{eq-24t1} was referred to in \cite{Ben-Dayan:2023inz}, as  an improved QNEC. However, the standard $G \to 0$ simply yields the previously known QNEC.} However, by dividing the RHS of Eq.~\eqref{eq-24t1} by $G$ and taking the $G \to 0$ limit, the ball-shaped region limit to the Rindler wedge (since the ball radius $R \sim \theta_\lambda^{-1} \sim G^{-1})$, and we obtain the QNEC as a pure QFT constraint. Explicitly, 
\begin{align}\label{eq-131q5}
    2\pi \langle T_{ij}\rangle k^i k^j -\frac{1}{\sqrt{h_\lambda}}\partial_\lambda \left.\frac{\delta S_{\rm ren}}{\delta V(y^a)}\right\rvert_\lambda \geq 0,
\end{align}
for $\mathcal{W}_\lambda$ null deformation of a Rindler wedge~\footnote{The QNEC can also be derived on an arbitrary curved background from the QFC~\cite{Bousso:2015mna} (and we expect also from the rQFC), subject to the local conditions $\theta_\lambda= \varsigma^2_\lambda=0$. Here, for simplicity we concentrate on the case of Minkowski space and the Rindler wedge, which in particular satisfies those conditions.}. Here, $\langle T_{ij}\rangle$ is being evaluate at $y^a$ on $\eth \mathcal{W}_\lambda$. \footnote{Though we only focus on the Rindler wedge case here, one can define QNEC in curved spacetime as well, and we expect that to follow from the rQFC as well. The $d>2$ QNEC in curved spacetime was proved for holographic CFTs in~\cite{Akers:2017ttv}.}

So far, we have shown that a standard $G \to 0$ limit of Eq.~\eqref{eq-24t1} leads to the QNEC. Let us now introduce a parameter $\Sigma$ which controls the transverse length scale of the null deformation profile $V_\lambda(y^a)$. For instance, we can take $V_\lambda(y^a)$ to be $\lambda$ times a bump function with characteristic width $\Sigma$:
\begin{align}
    V_\lambda(y^a)= \lambda~\exp\left[\frac{1}{1-\Sigma^2/ |y|^2}\right], ~~~~~~~|y| \leq \Sigma,
\end{align}
where $|y|$ denotes proper distance on $\eth \mathcal{W}$.

In Sec.~\ref{sect:CFT}, we find that a different $G \to 0$ limit of the rQFC, one in which we simultaneously send $\Sigma \to 0$ while keeping $\Sigma^{d-2} \gg G$ (as required for validity of perturbation theory) gives a new (stronger) bound on CFTs.\footnote{For instance, let $\Sigma = L^{1-\alpha} G^{\frac{\alpha}{d-2}}$ for $0<\alpha<1$, where $L$ is some large scale associated to the state under consideration.} This is because the limit creates a competition between the first and the second lines on the RHS of $\partial_\lambda \Theta_\lambda$ in Eq.~\eqref{eq-24t1}. We work this out explicitly in a class of near-vacuum states where we have a lot of computational control over the terms in Eq.~\eqref{eq-24t1}. We sketch the basic idea and main results in the rest of this section, and we also speculate about a more general stronger-than-QNEC bound.
\begin{figure}
    \centering
    \includegraphics[width=0.9\linewidth]{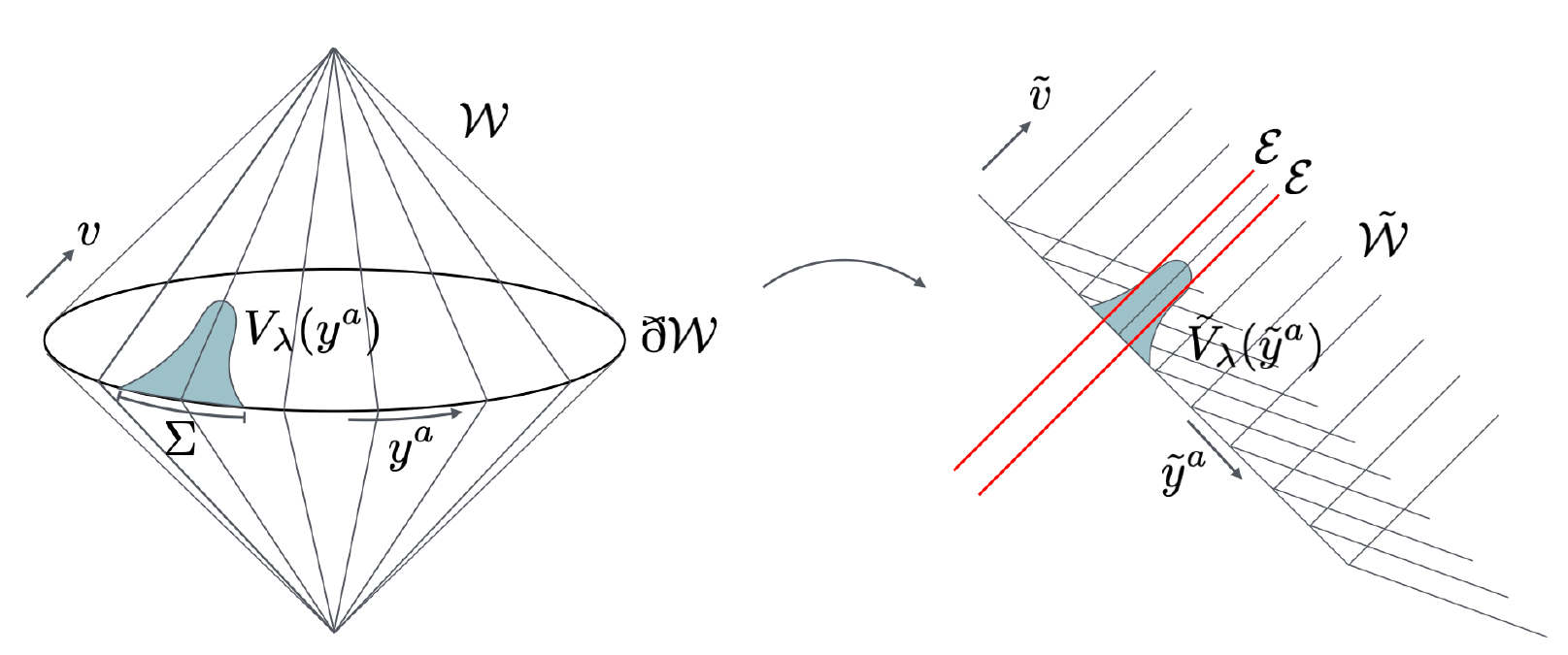}
    \caption{Application of the rQFC to the domain of dependence $\mathcal{W}$ of a ball in Minkowski space (left). We consider null deformations of the edge $\eth \mathcal{W}$, parametrized by $y^a$, to $v = V_\lambda(y^a)$ with width $\Sigma$. A new non-gravitational limit with $G, \Sigma \to 0$ and $\Sigma^{d-2} \gg G$ yields a stronger-than-QNEC bound in CFTs. As part of the derivation, $\mathcal{W}$ is conformally mapped to a Rindler wedge $\tilde{\mathcal{W}}$ with deformation $\tilde{v} = \tilde{V}_\lambda(\tilde{y}^a)$, where von Neumann entropy derivatives relate to the coincident limit of two $\mathcal{E}$ insertions.}
    \label{fig:placeholder}
\end{figure}
First, we use a conformal transformation to map the null-deformed ball-shaped regions $\mathcal{W}_\lambda$ to null-deformed Rindler wedges $\tilde{\mathcal{W}}_{\lambda}$. In a CFT, we can use this to relate the second line of Eq.~\eqref{eq-24t1} to its counter-part on the Rindler wedge. It turns out that:
\begin{align}\label{eq-fh9nnrgb}
    2\pi \langle T_{ij}\rangle k^i k^j -\frac{1}{\sqrt{h_\lambda}}\partial_\lambda \left.\frac{\delta S_{\rm ren}}{\delta V(y^a)}\right\rvert_\lambda - \frac{1}{\sqrt{h_\lambda}}\frac{2\theta_\lambda}{d-2} \left.\frac{\delta S_{\rm ren}}{\delta V(y^a)}\right\rvert_\lambda = 2\pi \langle \tilde{T}_{ij}\rangle \tilde{k}^i \tilde{k}^j -\frac{1}{\sqrt{\tilde{h}_\lambda}}\partial_{\lambda} \left.\frac{\delta \tilde{S}_{\rm ren}}{\delta \tilde{V}(\tilde{y}^a)}\right\rvert_{\lambda},
\end{align}
where the tilded quantities denote the analogous of the LHS quantities but pertaining to a Rindler wedge $\tilde{\mathcal{W}}$. In particular, $\tilde{y}^a$ denotes the transverse direction $\eth \tilde{\mathcal{W}}$, and $\tilde{k}^i = \partial_{\tilde{v}}$ denotes the affine null generators of $\partial^+\tilde{\mathcal{W}}$ (see Fig.\ref{fig:conformal_transform}). The wedge $\tilde{\mathcal{W}}_{\lambda}$ is then determined by the function $\tilde{v} = \tilde{V}(\tilde{y}^a)$. The (extra) third term on the LHS of Eq.~\eqref{eq-fh9nnrgb} comes from a non-linear relationship between $V_\lambda(y^a)$ and $\tilde{V}_{\lambda}(\tilde{y}^a)$.

We can combine Eq.~\eqref{eq-fh9nnrgb} with Eq.~\eqref{eq-24t1}, to obtain:
\begin{align}\label{eq-24t664}
    \Theta_\lambda(y^a)=0 \implies \partial_\lambda \Theta_\lambda(y^a) &=\frac{16(d-1)}{d-2} G^2 \left(\frac{1}{\sqrt{h_\lambda}}\left.\frac{\delta S_{\rm ren}}{\delta V(y^a)}\right\rvert_\lambda\right)^2\nonumber\\
    &- 4G \left(2\pi \langle \tilde{T}_{ij}\rangle \tilde{k}^i \tilde{k}^j -\frac{1}{\sqrt{\tilde{h}_\lambda}}\partial_{\lambda} \left.\frac{\delta \tilde{S}_{\rm ren}}{\delta \tilde{V}(\tilde{y}^a)}\right\rvert_{\lambda}\right) + \cdots.
\end{align}
It turns out that for a generic class of near-vacuum states:\footnote{The near-vacuum states are chosen for simplicity of the analysis in Sec.~\ref{sect:CFT} to have no stress-energy tensor in the vicinity of the deformation profile.}
\begin{align}\label{eq-nearvac}
    \ket{\psi} = \ket{0} + i \epsilon \ket{\chi} + O(\epsilon^2),
\end{align}
the Rindler wedge quantity on the second line of Eq.~\eqref{eq-24t664} can be computed explicitly~\cite{Balakrishnan:2019gxl} (see also~\cite{Leichenauer:2018obf}, and Appendix~\ref{sec:EE} for a review). In particular, a non-trivial scaling with $\Sigma$ as $\Sigma \to 0$ can be obtained:
\begin{align}\label{eq-1234}
    2\pi \langle \tilde{T}_{ij}\rangle \tilde{k}^i \tilde{k}^j -\frac{1}{\sqrt{\tilde{h}_\lambda}}\partial_{\lambda} \left.\frac{\delta \tilde{S}_{\rm ren}}{\delta \tilde{V}(\tilde{y}^a)}\right\rvert_{\lambda} = O(\epsilon^2\Sigma^{d-2-\delta})
\end{align}
where $\delta$ depends on both the theory and the state. We will discuss it in more detail below. Furthermore, we expect that generically:\footnote{Here, we use that in the $G\to0$ limit, $\delta S_{\rm ren}/\delta V(y^a)$ goes to its Rindler wedge counter-part denoted by $\delta \tilde{S}_{\rm ren}/\delta \tilde{V}(\tilde{y}^a)$.}
\begin{align}\label{eq-generic}
    \left.\frac{1}{\sqrt{\tilde{h}_\lambda}}\frac{\delta \tilde{S}_{\rm ren}}{\delta \tilde{V}(\tilde{y}^a)}\right\rvert_\lambda = O(\epsilon),
\end{align}
which implies that the $h_\lambda^{-1/2} \delta S_{\rm ren} / \delta V(y^a) = O(\epsilon)$.

Let us discuss how Eqs.~\eqref{eq-24t664},~\eqref{eq-1234}, and \eqref{eq-generic} result in a bound stronger than the QNEC. As explained, the regular $G\to 0$ limit of Eq.~\eqref{eq-24t664} yields the QNEC. However, Eq.~\eqref{eq-1234} in principle allows us to suppress the second line of Eq.~\eqref{eq-24t664} with an independent parameter $\Sigma$ compared to the \emph{positive} first line. Therefore, a simultaneous $G,\Sigma\to 0$ limit must extract a new bound from the rQFC.\footnote{In Sec.~\ref{sect:CFT} and Appendix~\ref{sec:Q}, we argue that the omitted terms (in the ellipsis) are $o(G^2)$ and therefore do not enter this analysis.} But since $\Sigma$ controls the transverse size of our deformation profile, to stay in the regime of validity of the computation, it is important that we obey:
\begin{align}\label{eq-EFT}
    \Sigma^{d-2} \gg G.
\end{align}
It is easy to see that if $\delta<0$ in Eq.~\eqref{eq-1234}, it is possible to make $\partial_\lambda \Theta_\lambda$ negative while staying within the regime~\eqref{eq-EFT}.

The conclusion is that an appropriate $G,\Sigma\to0$ limit of the rQFC implies the following stronger-than-QNEC bound in all CFTs:
\begin{align}\label{eq-newbound}
    \left.\frac{1}{\sqrt{\tilde{h}_\lambda}}\frac{\delta \tilde{S}_{\rm ren}}{\delta \tilde{V}(\tilde{y}^a)}\right\rvert_\lambda = O(\epsilon) \implies 2\pi \langle \tilde{T}_{ij}\rangle \tilde{k}^i \tilde{k}^j -\frac{1}{\sqrt{\tilde{h}_\lambda}}\partial_{\lambda} \left.\frac{\delta \tilde{S}_{\rm ren}}{\delta \tilde{V}(\tilde{y}^a)}\right\rvert_{\lambda} \stackrel{\Sigma\to0}\geq O(\epsilon^2\Sigma^{d-2})\geq0,
\end{align}
where by the notation $\geq O(\epsilon^2\Sigma^{d-2})$, we mean that at $O(\epsilon^2)$, the expression cannot go to zero faster than $\Sigma^{d-2}$ as $\Sigma\to0$. In particular, this bound is stronger than the QNEC applied to this state, which only demands that the LHS of Eq.~\eqref{eq-1234} is positive.

Let us now comment on the origin of the parameter $\delta$ in Eq.~\eqref{eq-1234}. In~\cite{Balakrishnan:2019gxl}, it was found that $\delta$ in Eq.\eqref{eq-1234} can be obtained from the following:
\begin{align}
    \int_{-\infty}^{\infty} ds \bra{\chi_s} \mathcal{E}(\tilde{y}^a) &\mathcal{E}(\tilde{y}'^a) \ket{\chi_s} \stackrel{\tilde{y}\to \tilde{y}'}= O(|\tilde{y} - \tilde{y}'|)^{-\delta}),\nonumber\\
    \text{where,}~~~\mathcal{E}(\tilde{y}^a) =& \int_{-\infty}^{\infty} d \tilde{v}~T_{ij}(\tilde{v}, \tilde{y}^a) \tilde{k}^i \tilde{k}^j.
\end{align}
Here, $|\tilde{y} - \tilde{y}'|$ denotes the proper distance between $\tilde{y}^a$ and $\tilde{y}'^a$, and
\begin{align}
    \ket{\chi_s} = e^{-i K s} \ket{\chi}
\end{align}
where $K$ is the boost generator around the Rindler wedge. In Appendix~\ref{sec:v1}, we analyze the rQFC implication that $\delta\geq0$ from the perspective of the operator product expansion of $\mathcal{E}$'s.

We can summarize the story so far in the following diagram:
\begin{center}
\includegraphics[scale = 1.5]{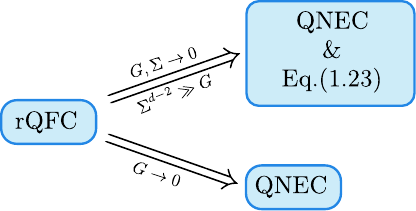}
\end{center}

Let us end the introduction with a speculation. Our new CFT bound \eqref{eq-newbound} is stronger than the QNEC in the particular class of near-vacuum states \eqref{eq-nearvac}. But it is unsatisfactory to only have such a bound in this limited regime. It is natural to expect that, in fact, a universal (holding in all QFTs in all states) stronger-than-QNEC bound is true. In the $\Sigma \to 0$ regime, this bound would presumably prevent the QNEC from scaling with $\Sigma^\alpha$ with $\alpha> d-2$, whenever $\delta S_{\rm ren} / \delta V(y^a) \neq 0$. These conditions can culminate in the following natural bound for null deformations of the Rindler wedge:
\begin{align}\label{eq_34t1wef}
    2\pi \langle T_{ij}\rangle k^i k^j -\frac{1}{\sqrt{h_\lambda}}\partial_\lambda \left.\frac{\delta S_{\rm ren}}{\delta V(y^a)}\right\rvert_\lambda \stackrel{\Sigma \to 0}\geq \kappa \Sigma^{d-2} \left(\frac{1}{\sqrt{h_\lambda}} \left.\frac{\delta S_{\rm ren}}{\delta V(y^a)}\right\rvert_\lambda\right)^2, ~~~~~~\text{(Speculative)}
\end{align}
where $\kappa>0$ is to be determined. This bound would then hold by either the LHS scaling with $\Sigma^\alpha$ as $\Sigma \to 0$ for $0\leq \alpha < d-2$, or with $\alpha = d-2$ in which case the competition of magnitudes between LHS and RHS becomes non-trivial. The bound~\eqref{eq_34t1wef} is reminiscent of the $d=2$ strengthened QNEC in CFTs, which was proposed in~\cite{Wall:2011kb}, and proved in general states in~\cite{Almheiri:2019yqk}. We leave an exploration of this bound to future work.

\section{Review of the semiclassical gravity regime}
\label{sect:semi}

In this section, we give a brief review of the semiclassical gravity regime which we employ in this paper. Since this is just an overview of basics, readers who feel confident about what is meant by the regime \eqref{eq-1c24156c}, and the resulting action, and generalized entropy functional, can skip ahead to the next sections. See~\cite{Parker:1992zp} and some references within for a more extensive discussion of the semiclassical gravity regime, including some technical challenges, and some of their proposed resolutions.

As mentioned in the introduction, our definitions of the quantum expansion $\Theta$ in Eq.~\eqref{eq-11cx5} (and therefore the QFC and rQFC) require the existence of a classical background metric. In perturbative quantum gravity, one starts with a classical background $g_{ij}$ and quantizes graviton fluctuations $\delta \hat{g}_{ij}$ (along with any matter fields) on that background. Therefore, backreaction is described by a $\delta \hat{g}_{ij}$ operator in general, not a classical field. In particular, the fluctuations, i.e., $\langle\delta \hat{g}_{ij}^2\rangle - \langle\delta \hat{g}_{ij}\rangle^2$, may be large, and therefore it is only at the zeroth order in $G$ expansion where we get to describe the geometry classically.

The need for a classical background to use our definition of $\Theta$ then forces us to consider only the leading in $G$ term in the generalized entropy functional~\eqref{eq-13r3r}, namely, the $A(\eth \mathcal{W})/4G$ term. In particular, it would be inconsistent to include, say a correction to $S_{\rm gen}$ such as the renormalized entropy term without also including an analogous $G$ correction to the metric which, as discussed, is described by an operator.\footnote{Another way to see why we cannot include say the renormalized entropy corrections to the generalized entropy of a general wedge is the absence of a satisfactory prescription for computing the von Neuman entropy of gravitons in an arbitrary wedge. This is an issue since, in perturbative quantum gravity, the graviton entropy can be of the same order as the matter entropy in general.} Forgoing the renormalized entropy contribution to $S_{\rm gen}$ would be a tremendous limitation because many non-trivial applications of semiclassical constraints like the rQFC are hidden in the renormalized entropy contribution (e.g., the generalized second law: $\Theta\geq 0$ on the horizon of an evaporating black hole due to Hawking radiation). At the same time, defining $\Theta$ including higher $G$ corrections would require extending the definition~\eqref{eq-11cx5} to incorporate notions of the wedge and the null direction in a (perturbatively) quantum geometry. While this is an interesting technical challenge, here we resort to a simpler way to get around this issue. It is feasible that our results can be extended without much modification to the perturbative quantum gravity regime once the quantum geometry hurdle is overcome.\footnote{For instance, in~\cite{Wall:2011hj}, the generalized second law for perturbations to a Killing horizon was proved in perturbative quantum gravity. A crucial step was to define the area of cuts of the horizon as an operator.}

In order to have a classical geometry on which the generalized entropy functional includes correction including the renormalized entropy term, a standard approach is to consider a \emph{semiclassical gravity} regime involving a large number $c$ of matter fields. The regime is defined as in Eq.~\eqref{eq-1c24156c} in the introduction which we reproduce here:
\begin{align}\label{eq-1c24156c2}
    G \to 0, ~~ c G \equiv \ell_S^{d-2} = \text{fixed}.
\end{align}
Equivalently, we can view this limit as $c\to \infty$, with the species scale $\ell_S$ held fixed. Therefore, when the matter species are excited together, we obtain $\langle T_{ij} \rangle \sim \langle T_{ij}^2 \rangle - \langle T_{ij} \rangle^2 \sim c$, therefore the relative fluctuations of the stress-energy tensor and in turn of the backreacted geometry go to zero in the $c \to \infty$ limit.\footnote{To see this, let us model the total stress tensor expectation value as the sum of independent random variables $x_1,\dots,x_c$ with 
$\mathbb{E}[x_a]=\mu\neq 0$ and $\mathrm{Var}(x_a)=\sigma^2<\infty$. 
For the sum $S_c=\sum_{a=1}^c x_a$ we have 
$\mathbb{E}[S_c]=c\mu$ and $\mathrm{Var}(S_c)=c\sigma^2$, 
so the relative fluctuations obey 
$\sqrt{\mathrm{Var}(S_c)}/\mathbb{E}[S_c]\sim c^{-1/2}$.}

Note that in this regime one obtains a \emph{species scale} $\ell_S$ which controls the backreaction of the quantum matter fields. To obtain a \emph{perturbative} semiclassical gravity regime, one must set the species scale $\ell_S$ much smaller than other length scales in the problem. We then expect the following effective action:
\begin{align}\label{eq-24rv12}
I_{\rm effective}[g_{ij}] =  \frac{1}{16 \pi G}\int d^d x \sqrt{-g} &\left(R + \ell_S^2 \left(\alpha_1 R^2 + \alpha_2 R_{ij}R^{ij}+ \alpha_3 R_{ijkl}R^{ijkl}\right) + \cdots \right)\nonumber\\
&~~~~~~~~~~~~~~~~~~~~~~~~~~~~~~~~~~~~+ \log Z_{\rm matter}[g_{ij}] +\cdots,
\end{align}
where $R_{ijkl}$, $R_{ij}$, $R$ denote the Riemann tensor, Ricci tensor, and scalar respectively, and $Z_{\rm matter}[g_{ij}]$ denotes the renormalized matter partition function on a spacetime with metric $g_{ij}$. The ellipsis on the first line denotes further higher curvature gravity terms which are suppressed by appropriate powers of $\ell_S$~\footnote{Note that are higher curvature corrections of this order are \emph{induced} by the matter sector anyways, so suppressing them further would be unnatural. In general, one may include bare higher curvature terms which to enlarge the magnitude of higher curvature corrections, but since nothing in our discussion changes with such terms, the minimal ansatz of Eq.~\eqref{eq-24rv12} suffices for us.}, and the ellipsis on the second line indicate any terms of order $c\ell_S^n$ for $n>0$. One could also add a cosmological constant term to the action \eqref{eq-24rv12}, though we have omitted it for simplicity. Also, $G$ and the $O(1)$ $\alpha_n$ denote the (renormalized) Newton's constant and Wilson coefficients respectively.\footnote{Systematically, one must start with bare couplings, matter fields action of the many fields, and a proper set of counter-terms, which then lead to renormalized parameters in the effective action \eqref{eq-24rv12}.}

Note that naturally, in the perturbative semiclassical regime (i.e., \eqref{eq-1c24156c2} with $\ell_S$ perturbatively small) physical quantities organize themselves, at leading order in $c$, in an all-orders expansion in $\ell_S$. This can become manifest in the action \eqref{eq-24rv12} by taking out a prefactor $c$ which multiplies an expansion in $\ell_S$. For example, the Einstein Hilbert term is $O(\ell_S^{2-d})$, while the $\log Z_{\rm matter}$ term is $O(\ell_S^0)$ in this expansion. The generalized entropy organizes itself in a similar manner:
\begin{align}\label{eq-XX}
    S_{\rm gen}(\mathcal{W}) = c \left[\frac{A(\eth \mathcal{W})}{4 \ell_S^{d-2}} + \cdots + S_{\rm ren}^{(1)}(\mathcal{W}) + o(\ell_S^0)\right],
\end{align}
where the leading term is the usual Bekenstein-Hawking area term (recall the relation $c G_d = \ell_S^{d-2}$), and $S_{\rm ren}^{(1)} = S_{\rm ren} / c$ is the renormalized von Neumann entropy per species in $\mathcal{W}$. Besides $S_{\rm ren}^{(1)}$, it is natural to lump in together all terms with non-positive powers of $\ell_S$, which we denote by $S_{\rm local} (\mathcal{W})$. Quite generically, $S_{\rm local} (\mathcal{W})$ is a sum of local geometric contributions:
\begin{align}\label{eq-1112231}
    S_{\rm local}(\mathcal{W}) = c \sum_{k=0}^{\lfloor \frac{d-2}{2}\rfloor} \frac{\mathcal{I}_k(\eth \mathcal{W})}{\ell_S^{d-2-2k}}  = \frac{A(\eth \mathcal{W})}{4 G} + O(c \ell_S^{4-d}),
\end{align}
where $\mathcal{I}_k(\eth \mathcal{W})$ are integrals of local (covariant) geometric data on $\eth{\mathcal{W}}$ (scalars made up of intrinsic, extrinsic, and ambient geometric tensors at $\eth \mathcal{W}$). For instance, $\mathcal{I}_0 = \int_{\eth \mathcal{W}} d^{d-2} y ~\sqrt{h}$. The higher order corrections, which are in correspondence with the higher curvature gravity corrections in the action \eqref{eq-24rv12}, are known as the Dong entropy terms (see~\cite{Myers:2013lva, Dong:2013qoa} for a prescription to compute them).\footnote{In even $d$, the last term in $S_{\rm local}$ is independent of $\ell_S$, and it can be viewed as getting renormalized by the logarithmic divergence (present in even $d$) of the regularized von Neumann entropy.} See Appendix~\ref{sec:Q} for an explicit example of a semiclassical gravity limit in the braneworld model where the action \eqref{eq-24rv12} and $S_{\rm gen}$ expansions can be explicitly derived.

Lastly, we denote by $Q$, terms in $S_{\rm gen}$ with positive powers of $\ell_S$. These contributions can be divided into local terms (which are the continuation of $S_{\rm local}$-type terms), and fully non-local terms. The terms $Q$ are not extensively discussed in the literature. In Sec.~\ref{sect:CFT}, we make an assumption about the order of magnitude of their contribution to $\partial_\lambda\Theta_\lambda$ which we give evidence for in Appendix~\ref{sec:Q}.

In the next two sections, we operate in the semiclassical gravity regime. In Sec.~\ref{sect:JT}, we work with a toy model that is simple enough that the generalized entropy simply equals the ``schematic form'' in Eq.~\eqref{eq-13r3r}. However, in Sec.~\ref{sect:CFT}, we discuss a general CFT in $d>2$ coupled to gravity and in the perturbative semiclassical regime. There, we follow closely the discussion of this section and work with the $S_{\rm gen}$ ansatz of Eq.~\eqref{eq-XX}. We do not believe that the employment of the semiclassical regime is the only way to make sense of the conclusion in these sections. It is merely a way to avoid the technical hurdles discussed at the beginning of this section. Generalizing the discussion beyond this limit (e.g., to the ordinary perturbative quantum gravity regime) will be left to future work.

\section{The rQFC and QFC in JT gravity coupled to a QFT}
\label{sect:JT}

In this section, we flash the basics of JT gravity (for a thorough review, see \eg,~\cite{Mertens:2022irh}). We then present the simple proof of the rQFC in JT gravity minimally coupled to a QFT. The strategy of the proof is similar to that of the proof of the quantum Bousso bound \cite{Strominger:2003br,Bousso:2015mna} in JT gravity presented in \cite{Franken:2023ugu}. In particular, we recycle a result by Almheiri, Mahajan, and Maldacena \cite{Almheiri:2019yqk}, rephrasing it as the rQFC. Finally, we present two JT gravity setups that exhibit explicit violations of the QFC.

\subsection*{JT gravity coupled to a QFT}

A fruitful direction to explore quantum gravity has been to study lower dimensional toy models. The two-dimensional Einstein-Hilbert action is topological, and the associated Einstein tensor vanishes. A well-known prescription to create a non-trivial two-dimensional gravity theory is to couple the Ricci tensor to a scalar field $\phi$, known as the dilaton. The most general two-derivative dilaton gravity action can always be reduced to
\begin{equation}
    I = \frac{1}{16\pi G}\int d^2x \sqrt{-g}(\phi R + U(\phi)),
\end{equation}
where $U(\phi)$ is the dilaton potential \cite{Mertens:2022irh}. Our special focus, Jackiw–Teitelboim (JT) gravity, is the quantum gravity toy model obtained by adopting a linear potential $U(\phi)= - \Lambda \phi$ \cite{Jackiw:1984je, Teitelboim:1983ux}, providing a solvable toy model of quantum gravity \cite{Henneaux:1985nw,Saad:2019lba}. When focusing on Anti-de Sitter and de Sitter space with radius of curvature $L_{\rm (A)dS}$, we set $\Lambda=-2/L_{\rm AdS}^2$ and $\Lambda=+2/L_{\rm dS}^2$, respectively. The complete JT gravity action on a two-dimensional manifold $\mathcal{M}$ reads \cite{Jackiw:1984je, Teitelboim:1983ux, Cotler:2019nbi}
\begin{equation}
\label{eq:JT}
    I_{\rm JT} = \frac{1}{16\pi G}\int_{\mathcal{M}} d^2x \sqrt{-g}((\phi_0+\phi) R -\Lambda \phi)+ \frac{1}{8\pi G}\int_{\partial\mathcal{M}} dx \sqrt{-h}(\phi_0K+\phi (K-1)),
\end{equation}
where we added the topological term $\phi_0 R$, the Gibbons-Hawking-York boundary term and the holographic counterterm ``$-1$'' \cite{York:1972sj,Gibbons:1976ue,Mertens:2022irh}. In Appendix \ref{sec:reduc}, we review how this action can be obtained from a dimensional reduction of the $d$-dimensional Einstein-Hilbert action (with $d\geq 3$) with cosmological constant. In this context, it is natural to define the total dilaton field:
\begin{equation}
    \Phi(x) = \phi_0+\phi(x).
\end{equation}
The JT action \eqref{eq:JT} can be supplemented with matter fields coupled to the metric. We will leave this QFT sector general unless otherwise specified, but as an example we can imagine the following matter sector consisting of $c$ massive scalar fields:
\begin{align}
\label{eq:matter}
    I_{\rm matter}[g_{ij}, \psi] = -\frac{1}{2}\sum_{n=1}^c\int d^2 x \sqrt{-g} ~(g^{ij }\nabla_i \psi_{(n)} \nabla_j \psi_{(n)} + m^2 \psi_{(n)}^2).
\end{align}
As in the introduction, we are interested in the limit,
\begin{align}\label{eq-1v1c4}
    G \to 0, ~~ \ell_S=c G = \text{fixed}.
\end{align}
In this limit, the quantum fluctuations of the dilaton can be ignored, and one obtains the semiclassical equations of motion:
\begin{equation}
\label{eq:dilaton_eom}
\left(g_{ij}\nabla^2-\nabla_{i}\nabla_{j}+g_{ij}\frac{\Lambda}{2}\right)\phi=8\pi G\Braket{T_{ij}},
\end{equation}
where $\langle T_{ij}\rangle$ denotes the expectation value of the stress-energy tensor. The only difference between AdS and dS JT gravity lies in the third term on the LHS of Eq.~\eqref{eq:dilaton_eom}, depending on the sign of $\Lambda$. Of particular importance for the discussion of quantum focusing is the component of Eq.~\eqref{eq:dilaton_eom} obtained by contracting with any null vector $k^{i}$:
\begin{equation}\label{eq:contracted_eom}
\frac{d^2\phi}{d{\lambda}^2}\equiv k^{i}k^{j}\nabla_{i}\nabla_{j}\phi=-8\pi G k^{i}k^{j}\Braket{T_{ij}}\equiv -8\pi G \Braket{T_{kk}},
\end{equation}
where we defined the derivative
\begin{equation}
    \frac{d}{d\lambda} = k^{\mu}\nabla_{\mu}.
\end{equation}

\subsection*{A proof of rQFC}

\begin{figure}[t]
    \centering
    \includegraphics[width=0.65\linewidth]{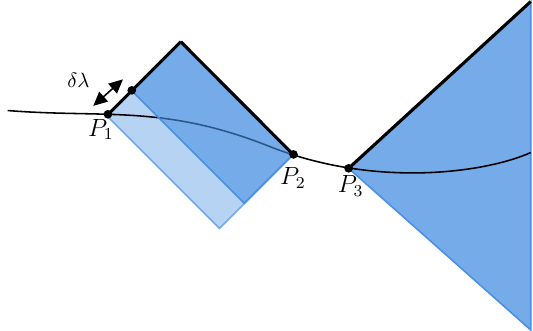}
    \caption{Let $\mathcal{W}$ be a wedge (light blue) whose edge is the union of points $P_i$. We consider variations of $S_{\rm gen}$ at $P_1$. The quantum expansion $\Theta\vert_{\mathcal{W}}(P_1)$ is defined as the variation of $S_{\rm gen}(\mathcal{W})$ with respect to $\lambda$, the affine parameter of a lightray $L^+(P_1)$ emanating from $P_1$ and included in $\partial^+\mathcal{W}$ (black lines). In this section we focus on $\Theta_{\lambda}=\Theta\vert_{\mathcal{W}_{\lambda}}$ with the $1$-parameter family of wedges $\mathcal{W}_{\lambda}$ defined as $\eth \mathcal{W_{\lambda}}= \tilde{P}(\lambda)\cup P_2\cup P_3$, where $\tilde{P}(\lambda)$ is a point on $L^+(P_1)$ at $\lambda$. The wedge $\mathcal{W}_{\lambda+\delta\lambda}$ is schematically pictured in dark blue. } 
    \label{fig:wedge}
\end{figure}

In JT gravity coupled to matter, we associate a generalized entropy to a spacetime wedge $\mathcal{W}$ using the following formula:
\begin{equation}
\label{eq:Sgen}
    S_{\rm gen}(\mathcal{W}) = \sum_i\frac{\Phi(P_i)}{4G} + S_{\rm ren}(\mathcal{W}).
\end{equation}
This definition is motivated by the transverse area interpretation of $\Phi$ (see Appendix \ref{sec:reduc} and Eq. \eqref{eq:JTarea}) and the fact that for any wedge $\mathcal{W}$ in $d=2$, we have:
\begin{align}
\eth \mathcal{W} =\bigcup_i P_i,
\end{align}
where $\{P_i\}$ is a discrete set of points. See Fig.~\ref{fig:wedge} for a schematic picture of the setup. 
The QFT entropy term $S_{\rm ren}(\mathcal{W})$ contains a universal logarithmic divergence, which is irrelevant for our purposes since we are only interested in differences of $S_{\rm gen}$ between wedges.

Let us comment on the origin of Eq.~\eqref{eq:Sgen} from the $d=2$ perspective. It is a cutoff-independent quantity, well defined in the perturbative regime, which is a non-trivial property that must be satisfied by the von Neumann entropy of subregions in semiclassical gravity and JT gravity \cite{Witten:2021unn, Penington:2023dql}. Moreover, the QFC which we are about to define is directly related to fundamental constraints on the algebraic structure of the theory, which yields a definition of entropy consistent with euclidean computations \cite{Penington:2023dql}. Finally, the generalized entropy \eqref{eq:Sgen} was shown to be identical to the Wald entropy, which is the Noether charge associated with a bifurcation point of a Killing horizon \cite{Pedraza:2021cvx}.\footnote{In two dimensions, any point of dS or AdS is the bifurcation point of a Killing horizon, so that a Wald entropy can be associated with any point.} Additionally, the authors of \cite{Pedraza:2021cvx} discussed the thermodynamics of generalized entropy, deriving a generalized first and second law.

Following Eq.~\eqref{eq-11cx5}, the quantum expansion at a given point in $\eth \mathcal{W}$ is given by:
\begin{equation}\label{eq-32cr11}
    \Theta_{\lambda} = \frac{4G}{\Phi}\frac{dS_{\rm gen}}{d\lambda},
\end{equation}
where $S_{\rm gen} = S_{\rm gen}(\mathcal{W}_{\lambda})$, with $\mathcal{W}_{\lambda}$ a $1$-parameter family of wedges parametrized by the affine parameter on the lightray emanating from one of the points in $\eth \mathcal{W_{\lambda}}$ (See Fig.~\ref{fig:wedge}). The division by $\Phi$ accounts for the factor $\sqrt{h}$ in Eq.~\eqref{eq-11cx5}, given the transverse area interpretation of $\Phi$ in this model.
In the classical limit, the quantum expansion $\Theta_{\lambda}$ reduces to the classical expansion $\theta_{\lambda}$:
\begin{equation}
    \Theta_{\lambda} \xrightarrow[\hbar\rightarrow 0]{}\theta_{\lambda} = \frac{1}{\Phi}\frac{d\Phi}{d\lambda}.
\end{equation}
This quantity has no direct interpretation in JT gravity seen as a genuine two-dimensional theory. However, it can be seen as the dimensional reduction of the expansion scalar in higher dimension. In particular, we review in Appendix \ref{sec:reduc} that it satisfies the classical focusing theorem under the null energy condition.

In this section, the only variable of interest will be $\lambda$. We will thus make the dependence on this variable implicit: $\Theta_{\lambda} = \Theta$, $\theta_{\lambda} = \theta$, and write $d/d\lambda=~'$. 

Next, to make contact with quantum focusing, we take an additional derivative of Eq.~\eqref{eq-32cr11} with respect to $\lambda$, obtaining:
\begin{equation}
\label{eq:dev}
    \Theta' = -8\pi G \frac{\Braket{T_{kk}}}{\Phi}-\theta^2+4G\left(\frac{S_{\rm ren}''}{\Phi}-\frac{\theta S_{\rm ren}'}{\Phi}\right),
\end{equation}
where we used the equations of motion \eqref{eq:contracted_eom}. The quantum null energy condition (QNEC) in two dimensions reads \cite{Bousso:2015mna}:
\begin{equation}
\label{eq:QNEC}
    2\pi\Braket{T_{kk}} \geq S_{\rm ren}'',
\end{equation}
and was proven in two-dimensional holographic QFT and CFT \cite{Koeller:2015qmn,Almheiri:2019yqk}.

Combined with \eqref{eq:dev}, and assuming $\Phi>0$,\footnote{In the regions where $\Phi<0$, one gets $\Theta'\geq -\theta\Theta$. While solutions with $\Phi<0$ are admissible when JT gravity is treated as a genuine two-dimensional theory, they are excluded if JT arises from dimensional reduction, where $\Phi>0$ by construction. An additional motivation to take $\Phi>0$ is that $G/\Phi$ may be interpreted as an effective coupling constant. We therefore find the restriction to $\Phi>0$ natural on physical grounds. 
} the QNEC \eqref{eq:QNEC} provides an upper bound on $\Theta'$:
\begin{equation}
\label{eq:lowerbound}
    \Theta' \leq -\theta \Theta.
\end{equation}
This inequality implies the restricted quantum focusing conjecture: for $\Theta=0$, one gets $\Theta'\leq 0$.

It is not clear whether this statement directly maps to higher dimensions. In particular, our derivation of this upper bound relies on the two-dimensional QNEC. This energy bound is not universally valid in semiclassical gravity. In particular, in $d>2$ spacetime dimensions, it only applies to null hypersurface with vanishing expansion \cite{Bousso:2015wca}, as explained in the introduction. We therefore do not expect the bound $\Theta'\leq -\theta\Theta$ to be related to an exact statement in semiclassical gravity in $d>2$ dimensions.

\subsection*{Necessary condition for QFC violations}
\label{sec:QFCviolations}

The lower bound \eqref{eq:lowerbound} provides a necessary condition for violations of the QFC in the physical spacetime regions where $\Phi>0$:
\begin{equation}
\label{eq:necessarycond}
    \Theta' > 0 \implies \theta\Theta < 0.
\end{equation}
Although this condition is not sufficient, it is restrictive enough to hint at where to look if one wants to find a violation of the QFC. The condition \eqref{eq:necessarycond} implies that a surface violating the QFC should lie on a lightsheet that is not a quantum lightsheet, or vice versa.\footnote{A lighsheet is a codimension-$1$ null hypersurface with $\theta\leq 0$. A quantum lighsheet is a codimension-$1$ null hypersurface with $\Theta\leq 0$.} This type of null hypersurface exists in backgrounds with strong quantum effects, such as an evaporating horizon.

In particular, for conformal matter, we expect violations to appear in regimes where $\ell_S\gtrsim \phi_0$ and $\ell_S\gtrsim \phi_r$. In a two-dimensional CFT, the QNEC \eqref{eq:QNEC} implies the stronger inequality $2\pi \Braket{T_{kk}} - S_{\rm ren}'' - \frac{6}{c}(S_{\rm ren}')^2 \geq 0$ \cite{Wall:2011kb}. Inserting this into $\Theta'$ leads to:
\begin{align} \label{eq:thetaprim} \Theta'\Phi^2 &\leq -\left(\Phi'+\frac{2\ell_S}{c}S_{\rm ren}'\right)^2 - \frac{24\ell_S}{c^2}(S_{\rm ren}')^2\left(\Phi-\frac{\ell_S}{6}\right). \end{align} 
QFC violations thus necessitate $\Phi < \ell_S/6$. Consider a conformal vacuum state taking the large $c$ limit. Then,
\begin{equation} \label{eq:phil_S} \Phi(x) = \phi_0 + \phi_r \mathcal{F}(x) + \ell_S \mathcal{G}(x), \end{equation} where $\phi_r\mathcal{F}(x)$ is the solution to the classical equations of motion without matter, and $\mathcal{G}(x)$ is the state-dependent contribution to $\Phi$.
Inserting \eqref{eq:phil_S} into \eqref{eq:thetaprim}, we find
\begin{equation} \Theta'\vert_{\frac{\ell_S}{\phi_0},\frac{\ell_S}{\phi_r}\rightarrow 0} \leq 0.
\label{lim}
\end{equation} 
QFC violations can thus appear only when treating JT gravity coupled to matter as a genuine two-dimensional theory (still imposing $G\rightarrow 0$, $c\rightarrow\infty$ at $\ell_S=cG$ fixed to ensure the semiclassical limit), allowing quantum matter effects to be large. On the other hand, Eq. \eqref{lim} guarantees the validity of the QFC in regimes where semiclassical JT gravity has a higher-dimensional interpretation:
\begin{equation} 
\phi_0 \gg \phi_r \gg \ell_S,
\end{equation}
which requires matter quantum contributions to be small compared to classical terms ($\phi_r\gg \ell_S$), and that the dynamical dilaton field corresponds to a perturbation around an extremal black hole ($\phi_0\gg\phi_r$). See Appendix \ref{sec:reduc}.\footnote{Note that while $\phi_0\gg\phi_r$ is enough to treat the gravitational part of the action as a spherical reduction of Einstein gravity, the two-dimensional matter content is not derived from a higher-dimensional model, such that semiclassical JT gravity cannot be treated as dimensional reduction of semiclassical gravity in the strictest sense.}

In the remainder of this section, we present two examples of semiclassical JT gravity setups that violate the QFC. We use units where $L_{\rm (A)dS}=1$.

\subsection*{AdS$_2$ black hole in equilibrium with a bath}

\begin{figure}[ht]
    \centering
    \includegraphics[width=0.5\linewidth]{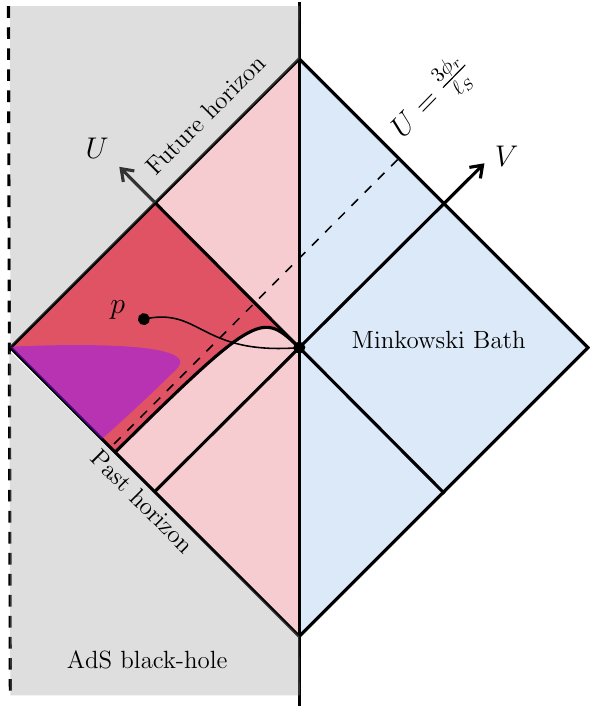}
    \caption{Penrose diagram of a two-dimensional extremal black hole in equilibrium with a bath at zero temperature. The blue region is half of Minkowski space and the red region is the Poincaré patch of the black hole. The future and past horizons are located at $U\rightarrow\infty$ and $V\rightarrow -\infty$ and the boundary of AdS is located at $U=V$. A spacelike slice with free endpoint $P$ and fixed endpoint at $(0,0)$ is depicted. The region where $\theta\Theta <0$ associated with $P$ along the $V$ direction, bounded by the curve which saturates the inequality \eqref{eq:bdy}, is depicted in dark red. In the limit $\phi_r\gg \ell_S$, it shrinks to a narrow strip along the future horizon and the null line $V=0$. The QFC violation region is shaded in purple.}
    \label{fig:AdS}
\end{figure}

Following \cite{Almheiri:2019psf,Almheiri:2019yqk}, we first consider an extremal black hole in AdS$_2$, in equilibrium with a bath modeling the asymptotically flat region far away from the black hole where gravity can be neglected. The two pieces of this toy model are constructed in semiclassical JT gravity, with one half of two-dimensional Minkowski space glued to the Poincaré patch of an extremal AdS$_2$ black hole along its conformal boundary, see Fig.~\ref{fig:AdS}.

The Minkowski half-space is described in null coordinates $(u,v)$ with $v>u$, while the black hole side is described by Poincaré coordinates $(U,V)$ where $U>V$, with metric:
\begin{equation}
    ds^2 = -\frac{4dUdV}{(U-V)^2}.
\end{equation}
We consider the black hole to be in the Poincaré vacuum $T_{UV}=T_{UU}=T_{VV}=0$ \cite{Spradlin:1999bn}. The bath is taken to be at zero temperature as it is in equilibrium with the extremal black hole, such that $T_{uv}=T_{uu}=T_{vv}=0$. Transparent boundary conditions imply that $(u,v)=(U,V)$ with a unique vanishing stress-energy tensor \cite{Almheiri:2019yqk}. The solution to the dilaton equations of motion \eqref{eq:contracted_eom} in the gravity region $U>V$ is:
\begin{equation}
\label{eq:dilAdS}
    \Phi = \phi_0 + \frac{2\phi_r}{U-V}.
\end{equation}
In \eqref{eq:Sgen}, we approximated the gravitational part of the entropy to be given by the area term. To ensure its validity, we impose $U-V\ll \phi_r$ \cite{Maldacena:2019cbz}. Beyond this regime, the gravitation theory becomes strongly coupled with a divergent effective gravitational constant ($G/\Phi$), and the bulk does not have a simple description. The entanglement entropy of a wedge $\mathcal{W}$, whose edge $\eth \mathcal{W}$ is the union of a point $P=(U,V)$ in AdS and a point $A=(U_A,V_A)$ in the Minkowski bath, is given by:
\begin{equation}
\label{eq:SAdS}
    S_{\rm ren}(\mathcal{W}) = \frac{c}{12}\log\left[\frac{(U-U_A)^2(V-V_A)^2}{(U-V)^2}\right] + {\rm constant}.
\end{equation}
The term $(U-V)^2$ in the denominator comes from the warp factor of the AdS$_2$ metric at the point $P$, and the constant term contains the bath UV cutoff. From now on, we fix $A$ at the origin, $(U_A,V_A)=(0,0)$, and let $P$ free in the AdS region spacelike separated to $A$, \ie~in the region $U>0$, $V<0$. We denote by $\theta$ and $\Theta$ the classical and quantum expansions associated with such point $P$, respectively.

As described in Eq.~\eqref{eq:necessarycond}, spacetime regions where violations of the QFC might happen can be probed by investigating the sign of the quantity $\theta\Theta$. Let us consider a light-ray at fixed $U$, along the null direction $V$, in the AdS region $U>V$ . Its classical expansion $\theta=\Phi'/\Phi$ satisfies $\theta>0$, as can be seen from Eq.~\eqref{eq:dilAdS}. If $U<3\phi_r/\ell_S$, then $\Theta>0$ for
\begin{equation}
\label{eq:bdy}
    V<-\frac{U^2}{\frac{3\phi_r}{\ell_S}-U},
\end{equation}
and $\Theta<0$ otherwise. If $U>3\phi_r/\ell_S$, then $\Theta<0$ for any $V<0$ . The region where $\theta\Theta<0$ is shown in dark red in Fig.~\ref{fig:AdS}.
In the limit $\phi_r\gg \ell_S$, it shrinks to a narrow strip along the future horizon and the null line $V=0$.

\begin{figure}[ht]
    \centering
    \includegraphics[width=0.8\linewidth]{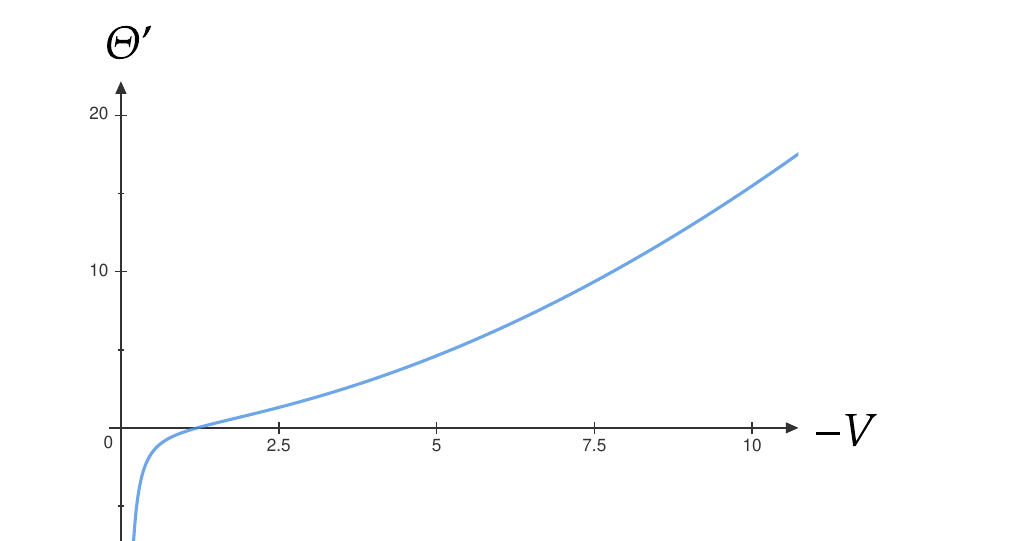}
    \caption{Plot of $\Theta'$ at fixed $U=1$, along the $-V$ direction. We have taken the constants $\phi_0=1,\phi_r=100,\ell_S=1000$.}
    \label{fig:AdSplot}
\end{figure}

We find QFC violations in the region $\theta\Theta < 0$, for a large range of parameters satisfying $\ell_S\gtrsim {\rm max}(\phi_0,\phi_r)$ , without restriction on the ratio $\phi_0/\phi_r$. We provide a schematic picture of the violation region in purple in Fig.~\ref{fig:AdS}, and plot the function $\Theta'$ along the null direction $-V$ at fixed $U=1$ in Fig.~\ref{fig:AdSplot}.
The analysis for the classical and quantum expansion is analogous for lightrays along $U$, as the expressions \eqref{eq:dilAdS} and \eqref{eq:SAdS} are symmetric around $U=-V$.

One can also consider a black hole at finite temperature. In this case, we find that QFC violations require a specific choice of Polyakov cosmological constant in the JT gravity action.\footnote{When considering $I_{\rm matter}$ to be a classical CFT, semiclassical backreaction is described by the Polyakov action \cite{Polyakov:1981rd} which in general includes a cosmological constant term proportional to $\lambda\Lambda$, where $\lambda$ is a parameter \cite{Fabbri:2005mw}. We find QFC violations for an unconventional gauge choice $\lambda=1$, which eliminates semiclassical corrections to the dilaton.}



\subsection*{Evaporating de Sitter horizon}
\label{sect:UdS}
Our second example of QFC violation is obtained by considering an evaporating de Sitter cosmological horizon, modeled in semiclassical JT gravity with positive cosmological constant. We are interested in the so-called full reduction model, characterized by a non-vanishing topological term $\phi_0>0$ as written in the action \eqref{eq:JT}. The geometry contains both a black hole and cosmological horizon, surrounding the static patch of a freely falling observer (see \eg \cite{Svesko:2022txo} for a review of the full reduction model of de Sitter JT gravity).

The Kruskal coordinates $(x^+,x^-)$ cover the static patches of two antipodal observers together with the expanding region of the de Sitter geometry behind their cosmological horizons. The metric in Kruskal coordinates is given by:
\begin{equation}
d s^2=-\frac{4}{(1-x^+x^-)^2}d x^+ d x^-.
\end{equation}
Another relevant set of coordinates are the static null coordinates $(\sigma^+,\sigma^-)$ covering the static patch associated with the south pole observer and related to the Kruskal coordinates by $x^{\pm}=\pm e^{\pm\sigma^{\pm}}$.
The metric in null static coordinates is given by:
\begin{equation}
    d s^2 = -\frac{1}{\cosh^2\left(\frac{\sigma^+-\sigma^-}{2}\right)} d\sigma^+ d\sigma^-.
\end{equation}
The most common choice of vacuum state for the conformal matter is the \emph{Bunch-Davies} state \cite{Bunch:1978yq}, obtained when the static patch is in thermal equilibrium with the exterior region, namely when there is equal ingoing and outgoing radiation in the static patch. It is analogous to the Minkowski vacuum \cite{PhysRevD.7.2850,Davies:1974th,Unruh} in flat space or the Hartle-Hawking state \cite{HH} of a black hole in thermal equilibrium with its Hawking radiation. In the Bunch-Davies vacuum, $T_{\pm\pm}^{(x)}(x^{\pm})=0$, while $T_{\pm\pm}^{(x)}(\sigma^{\pm})=c/(48\pi)$. This vacuum state is seen from a static observer as a thermal state at temperature $\beta^{-1}=1/2\pi$. Another vacuum of interest is the \emph{static vacuum}, which characterizes the state of a static observer in the $\sigma^{\pm}$ coordinates. It is analogous to the Rindler vacuum in flat space or the Boulware vacuum of a black hole \cite{Boulware,Spradlin:1999bn}, and obtained by considering $T_{\pm\pm}^{(\sigma)}(\sigma^{\pm})=0$, while $T_{\pm\pm}^{(\sigma)}(x^{\pm})=-c/(48 \pi (x^{\pm})^2)$. This state is seen as empty by a static observer following the trajectory $\sigma^+=\sigma^-$. 

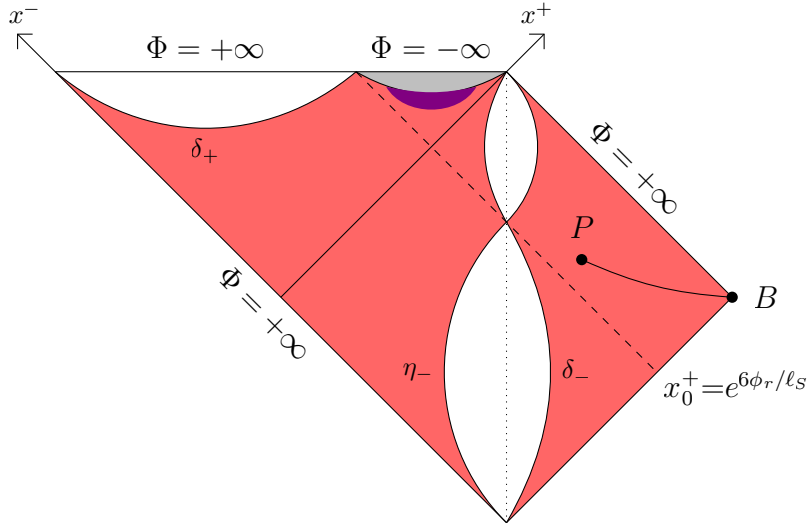
\begin{figure}[ht]
\centering
\begin{tikzpicture}
\begin{scope}[transparency group]
\begin{scope}[blend mode=multiply]
\path
       +(6,3)  coordinate (IItopright)
       +(0,3)  coordinate (IItopcenter)
       +(-6,3) coordinate (IItopleft)
       +(0,3)  coordinate (IIbotcenter)
       +(6,-3) coordinate (IIbotright)
       +(-6,-3) coordinate(IIbotleft)
      
       ;

\fill[fill=red!60] (IItopleft) to [bend right=40] (-2,3) to [bend right=30] (0,3) -- (-3,0) -- cycle;
\fill[fill=red!60] (-3,0) to (0,3) to [bend right=30] (0,1) to [bend right=45] (0,-3) -- (-3,0) -- cycle;
\fill[fill=red!60] (0,3) to[bend left=45] (0,1) to[bend left=30] (0,-3) to (3,0) to (0,3) -- cycle;

\fill[fill=gray!50] (-2,3) to[bend right=30] (0,3) to (-2,3);

\draw (-6,3) -- node[midway, above, sloped] {{$\Phi=+\infty$}} (-2,3) -- node[midway, above, sloped] {{$\Phi=-\infty$}} (0,3) ;

\draw[dotted] (0,3)--(0,-3);

\draw (0,3) -- node[midway, above, sloped] {{$\Phi=+\infty$}} (3,0) -- (0,-3) ;
\draw (-6,3) -- node[midway, below, sloped] {{$\Phi=+\infty$}} (0,-3) ;
\draw (-6.5,3.5) -- (-6,3) ;
\draw (-3,0) -- (0.5,3.5) ;

\draw (-6,3) to[bend right=40] node[midway, below, sloped] {{\footnotesize $\delta_+$}}(-2,3);

\draw (-2,3) to[bend right=30] (0,3);

\draw (0,1) to[bend left=30] node[midway, right] {{\footnotesize $\delta_-$}}(0,-3);

\draw (0,1) to[bend right=45] node[midway, left] {{\footnotesize $\eta_-$}}(0,-3);

\draw (0,3) to[bend left=45] (0,1);

\draw (0,3) to[bend right=30] (0,1);

\draw[dashed] (-2,3)--
(2,-1);

\draw (-6.5,3.3) -- (-6.5,3.5) -- node[midway, above, sloped] {{\footnotesize$x^-$}} (-6.3,3.5) ;
\draw (0.5,3.3) -- (0.5,3.5) -- node[midway, above, sloped] {{\footnotesize$x^+$}} (0.3,3.5) ;

\node at (1.8,-1.2) [label=right:$x_0^+{=}e^{6 \phi_r/\ell_S}$]{};

\draw (1,0.5) to[bend right=10] (3,0);

\node at (1,0.5) [label = above:{$P$}]{};
\node at (1,0.5) [circle, fill, inner sep=1.5 pt]{};

\node at (3,0) [label = right:{$B$}]{};
\node at (3,0) [circle, fill, inner sep=1.5 pt]{};

\end{scope}
\end{scope}

\fill[fill=violet] (-1.6,2.8) to[bend right=12] (-0.4,2.8) to[bend left=60] (-1.6,2.8) -- cycle;

\end{tikzpicture}
    \caption{ Penrose diagram for the backreacted geometry in the Unruh-de Sitter vacuum. The trajectory of a freely falling observer corresponds to the vertical dotted line. The region where $\Phi<0$ is gray shaded. Considering a light-ray along $x^+$, the region where $\theta \Theta <0$ is depicted in red. In the limit $\phi_r\gg \ell_S$, $x_0^+ \rightarrow +\infty$ while the curves $\delta_+$ and $\eta_-$ get closer to the cosmological horizons $x^{\pm}=0$. In this regime, the red shaded region thus reduces to narrow strips along the horizons. The QFC violation region is shaded in purple. A spacelike slice between a point $P$ in the bulk and the point $B$ at spatial infinity is also depicted.}
    \label{fig:Penrose_diag_UdS}
\end{figure}

Inspired by evaporating black holes, a vacuum state describing the evaporation of the de Sitter cosmological horizon can also be considered, in analogy with the Unruh vacuum of a black hole: the so-called \emph{Unruh-de Sitter} state \cite{Aalsma:2019rpt,Spradlin:1999bn,Unruh}. Such evaporation occurs when there is a positive net incoming energy flux on the static patch. Therefore, the Unruh-de Sitter state is obtained by considering asymmetrical vacuum states for the left and right-moving modes. The easiest possibility is to build a hybrid state of the Bunch-Davies and static de Sitter vacua, for the incoming and outgoing modes respectively. Considering the static patch with $x^+x^-\leq 0$, such state is thus defined by:
\begin{align}\label{eq:t_unruh_vac_1}
T_{++}^{(\sigma)}(x^{+})&=-\frac{c}{48\pi(x^{+})^2}, & T_{++}^{(\sigma)}(\sigma^{+})&=0,
\\
\label{eq:t_unruh_vac_2}
T_{--}^{(x)}(x^{-})&=0, & T_{--}^{(x)}(\sigma^{-})&=\frac{c}{48\pi}.
\end{align}
The equations of motion of the dilaton \eqref{eq:dilaton_eom} are then solved by
\begin{equation}\label{eq:sol_Unruh_vac}
    \Phi_{\rm UdS}(x^+,x^-) = \phi_0+ \phi_r\frac{1+x^+x^-}{1-x^+x^-} + \frac{\ell_S}{6}\left(1 - \frac{1+x^+x^-}{1-x^+x^-}\log x^+\right),
\end{equation}
with $x^+>0$. The dilaton is therefore divergent on the past cosmological horizon $x^+=0$ of an observer at the south pole, and on its future black hole horizon $x^+=+\infty$. It diverges to $+\infty$ and $-\infty$ at future infinity for $x<x_0^+$ or $x>x_0^+$ respectively, where $x_0^+\equiv e^{6 \phi_r/\ell_S}$. The Penrose diagram of the backreacted geometry is shown in Fig.~\ref{fig:Penrose_diag_UdS}.

The sign of the backreacted classical expansion $\theta$ of a given light-ray in the $x^+$ direction can be easily obtained by computing $\partial_+ \Phi_{\rm UdS}$. One finds that $\partial_+ \Phi_{\rm UdS} \leq 0$ for $\delta_-(x^+)<x^-<\delta_+(x^+)$, where $\delta_{\pm}(x^+)$ are the curves parameterized by: 
\begin{equation}
\delta_{\pm}(x^+)=\frac{1}{x^+}\left(\log\frac{x^+}{x_0^+}\pm\sqrt{1+\left(\log\frac{x^+}{x_0^+}\right)^2}\right).
\end{equation}

The entanglement entropy of a wedge $\mathcal{W}$, whose edge is defined as the union of $P=(x^+,x^-)$ and $B=(x_B^+,x_B^-)$, is given by \cite{Aalsma:2021bit}:
\begin{equation}
S_{\rm ren}(\mathcal{W})=\frac{c}{12}\log\left[
\frac{x^+ x_B^+ ~(x^- - x_B^-)^2~[\log(x_B^+/x^+)]^2}{(1-x^+ x^-)^2(1-x_B^+ x_B^-)^2}\right] + {\rm constant}.
\end{equation}
$B$ is arbitrary and will be sent to spatial infinity at the end of the computation, see Fig.~\ref{fig:Penrose_diag_UdS}. The sign of the quantum expansion $\Theta$ of a light-ray in the $x^+$ direction is obtained by computing $\partial_+ S_{\rm gen}$. After sending $x_B^{\pm}\rightarrow\pm\infty$, one finds that $\partial_+ S_{\rm gen}\geq 0$ for $\eta_-(x^+)<x^-<\eta_+(x^+)$, where $\eta_{\pm}(x^+)$ are the curves parameterized by:
\begin{equation}
\eta_{\pm}(x^+)=\frac{1}{x^+}\left(-\log\frac{x^+}{x_0^+}\pm\sqrt{1+\left(\log\frac{x^+}{x_0^+}\right)^2}\right).
\end{equation}
The curves $\delta_{\pm}$ and $\eta_-$ are drawn on the Penrose diagram of Fig.~\ref{fig:Penrose_diag_UdS}, while the curve $\eta_+$ lies in the region where $\Phi_{\rm UdS}<0$, depicted in gray. The regions where $\theta \Theta <0$ along $x^+$ are shaded in red. They correspond to the regions where violations of the QFC along $x^+$ can occur. In the limit $\phi_r\gg \ell_S$, $x_0^+$ goes to $+\infty$ while the curves $\delta_+$ and $\eta_-$ get closer to the cosmological horizons $x^{\pm}=0$. In this regime, the regions where violations of the QFC could occur therefore consist of narrow strips along the horizons. 

\begin{figure}[ht]
    \centering
    \includegraphics[width=0.7\linewidth]{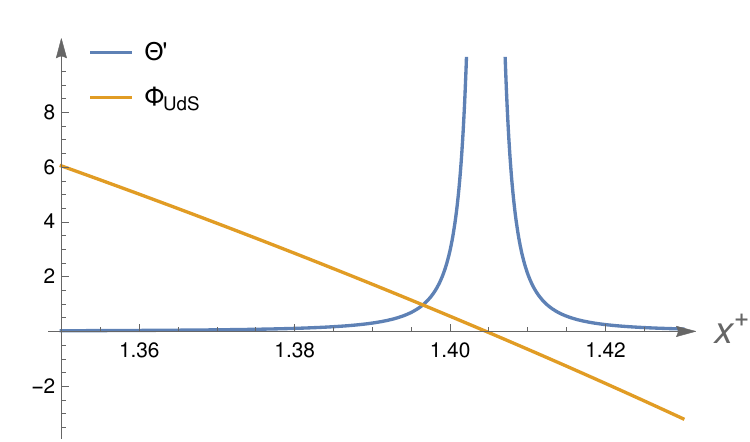}
    \caption{Plot of $\Theta'$ and $\Phi_{\rm UdS}$ at fixed $x^-=0.5$, along the $x^+$ direction. We have taken the constants $\phi_0=10,\phi_r=1,\ell_S=100$.}
    \label{fig:UdSplot}
\end{figure}

In the regime $\ell_S\gtrsim {\rm max}(\phi_0,\phi_r)$, independently of the ratio $\phi_0/\phi_r$, violations of the QFC along $x^+$ are found. We schematically depict the violation region in this regime in purple in Fig.~\ref{fig:Penrose_diag_UdS}, and plot the functions $\Theta'$ and $\Phi_{\rm UdS}$ along the null direction $x^+$ at fixed $x^-=0.5$ in Fig.~\ref{fig:UdSplot}.

In order to maintain contact with the higher-dimensional origin of JT gravity, the region where $\Phi_{\rm UdS}<0$ is not considered physical and should be removed from the spacetime. However, QFC violations occur even in the physical region where $\Phi_{\rm UdS}>0$.

\section{New \texorpdfstring{\boldmath $d>2$}{d>2} CFT bound from the rQFC}
\label{sect:CFT}

In this section, we test the rQFC in a particular setting in $d>2$ perturbative semiclassical gravity regime. We discover that the rQFC implies a novel stronger-than-QNEC bound in all CFTs.\footnote{The reader may wonder why instead of all CFTs, we do not demand the new constraint only of CFTs which arise as low energy limits of consistent UV-complete theories of quantum gravity. Historically, similar gravitational constraints, like the generalized second law (GSL), end up implying things about the low energy QFT sector (monotonicity of relative entropy in the case of the GSL) which are true in general QFT~\cite{Wall:2011hj}. So demanding the validity of the new constraint in all CFTs is informed by this precedent.} To obtain this result, we make crucial use of~\cite{Balakrishnan:2019gxl} which relates von Neumann entropy of null deformed Rindler wedges to light-ray operators. We end by speculating about a more universal strengthened QNEC.

\begin{figure}
    \centering
    \includegraphics[width=0.9\linewidth]{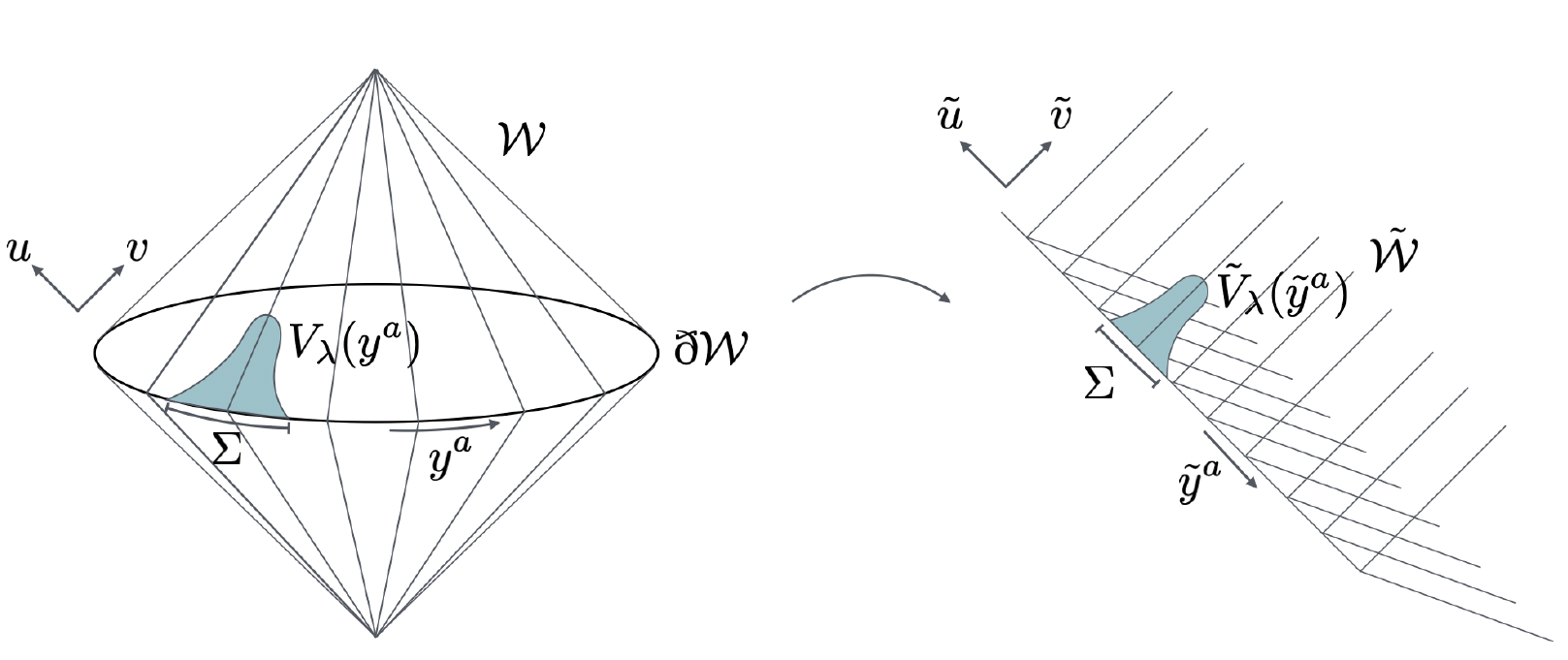}
    \caption{On the left, the domain of dependence of a ball-shaped region $\mathcal{W}$ is shown. By deforming its edge $\eth \mathcal{W}$ to $v = V_\lambda(y^a)$, where $y^a$ parametrizes $\eth \mathcal{W}$, we obtain a family of wedges. Here, $\Sigma$ denotes the width of $V_\lambda(y^a)$. Our argument involves a conformal transformation which maps $\mathcal{W}$ to a Rindler wedge $\tilde{\mathcal{W}}$, the region $(\tilde{u}<0, \tilde{v}>0, \tilde{y}^a)$, shown on the right. In the small $\Sigma$ limit, the deformation of $\mathcal{W}$, maps to a null deformation of the Rindler wedge parameterized by $\tilde{v} = \tilde{V}_{\lambda}(\tilde{y}^a)$, with characteristic width $\Sigma$, and related to $V_\lambda(y^a)$ through Eq.~\eqref{eq-1v256h}.}
    \label{fig:conformal_transform}
\end{figure}

\begin{figure}
    \centering
    \includegraphics[width=0.8\linewidth]{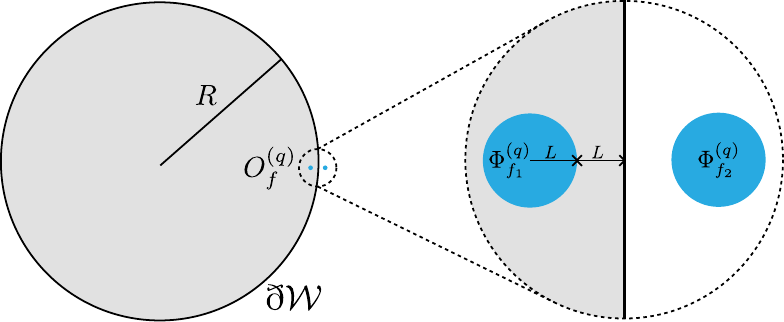}
    \caption{A timeslice of Minkowski spacetime is shown on which our ball-shaped region with edge $\eth \mathcal{\mathcal{W}}$ and radius $R$ lives. A simplifying aspect of our setup is to consider a near-vacuum state by acting on the vacuum by operator $\exp[i \epsilon \sum_{q=1}^c \Phi^{(q)}_{f_1} \Phi^{(q)}_{f_2}]$, where $\Phi^{(q)}_{f_1}$ and $\Phi^{(q)}_{f_2}$ denote smeared local Hermitian operators of matter species $j$, with smearing length scale $L\ll R$, and at a distance $L$ from $\eth \mathcal{W}$, and such that one operator is just inside and the other is just outside of the region. This operator does not inject any energy on $\eth \mathcal{W}$, allowing us to maintain the flat geometry and exact spherical shape of our region in the vicinity of the null deformation. The operator will non-trivially change the von Neumann entropy derivatives, allowing non-trivial tests of the rQFC.}
    \label{fig:nearvaccuum}
\end{figure}

Consider $c$ identical CFTs in the Minkowski vacuum in $d > 2$, and let $\mathcal{W}$, be the domain of dependence of a spatial ball with radius $R$. Let us work with the following coordinates in a neighborhood of $\eth\mathcal{W}$ (which is located at $u=v=0$):
\begin{align}\label{eq-4t1v241}
    ds^2 = -du\, dv + R^2 \left(1+\frac{v-u}{2R}\right)^2 d\Omega_{d-2}^2,
\end{align}
where $d\Omega_{d-2}^2$ is the metric of a unit $(d-2)$-sphere. Below, we will refer to these transverse coordinates collectively as $y^a$, but also often single out an angle $0\leq \chi \leq \pi$ among them with the following:
\begin{align}
    d\Omega_{d-2}^2 = d\chi^2 + \sin^2\chi~d\Omega_{d-3}^2.
\end{align}
By translating $\eth \mathcal{W}$ along $\partial^+\mathcal{W}$ by an affine length $v=V(y^a)$ we can obtain another wedge. Let us pick the following one-parameter family of such wedges labeled by $\lambda$:
\begin{align}
V_\lambda(y^a) = \lambda \exp \left(\frac{1}{1-\Sigma^2/R^2 \chi^2}\right), ~~|\chi|\leq \frac{\Sigma}{R}\label{eq-31455},
\end{align}
and $V_\lambda(y^a) = 0$ elsewhere. Therefore, the parameter $\Sigma$ determines the characteristic transverse width of the deformation profile (see left Fig.~\ref{fig:conformal_transform}). As we will see later in the section, dialing this width appropriately small will expose new implications of rQFC.
We take the perturbative semiclassical limit by sending $c \to \infty$ (or equivalently $G\to 0$), while holding the species scale $\ell_S = (cG)^{1/d-2}$ fixed. We further take $\ell_S$ to be perturbatively small compared to the physical scales of the problem (e.g., the size of the ball and the scale associated with the state). For a non-trivial test of rQFC, let us consider a ``near-vacuum'' state $\ket{\Psi}$ by acting on the vacuum with a unitary operator:
\begin{align}\label{eq-112131}
    \ket{\Psi} = \exp (i \epsilon \sum_{q=1}^c O^{(q)}_f) \ket{0},
\end{align}
where $\epsilon \ll 1$ and $O^{(q)}_f$ denotes a Hermitian operator in sector $q$ smeared by some compact smearing function $f$ (taken to be a scalar for simplicity). Explicitly,
\begin{align}
\label{eq:O}
    O^{(q)}_{f} = \int d^dx~f(x) O^{(q)}(x). 
\end{align}
To keep the analysis simpler, we will choose $O^{(q)}_f$ such that the entropy derivatives change non-trivially (see Eq.~\eqref{eq-XX}), but the geometry in a neighborhood of the null deformation (Eq.~\eqref{eq-31455}) remains unchanged. This can happen if the support of $O$ stays away from this region. For example, we can pick:
\begin{align}\label{eq-1c1311}
    O^{(q)}_f = \Phi^{(q)}_{f_1} \Phi^{(q)}_{f_2},
\end{align}
where smearing functions $f_1$ and $f_2$ are chosen to have compact support on a ball-shaped spacetime region with radius $L \ll R$, and is placed a spatial distance $L$ on the left and right side of $\eth \mathcal{W}$ around the locus of null deformation (see Fig.~\ref{fig:nearvaccuum}). Since this operator is unitary and is placed at spacelike separation to the locus of deformation, it does not inject any energy there. Note that this operator needs to be properly gravitationally dressed to be well-defined. But we are free to choose a dressing which does not change the geometry in a neighborhood of the null deformation, so to all orders in the $\ell_S$ perturbation theory the geometry is flat and:
\begin{align}\label{eq-2314v25v}
    \theta_\lambda (\chi=0) = \frac{d-2}{2R}, 
\end{align}
where $\theta_\lambda$ is the expansion of the congruence \eqref{eq-31455}.

Recall from Sec.\ref{sect:semi}, that in the perturbative semiclassical regime the generalized entropy can be written as an expansion in $\ell_S$:
\begin{align}\label{eq-24tr114t}
    S_{\rm gen}(\mathcal{W}) = c \left[\frac{A(\eth \mathcal{W})}{4 \ell_S^{d-2}} + \cdots + S_{\rm ren}^{(1)}(\mathcal{W}) + Q\right],
\end{align}
where $S^{(1)}_{\rm ren}= S_{\rm ren}/c$, is the renormalized entropy per species, and all the terms before $S_{\rm ren}$ are the geometrically local terms resulting from higher curvature corrections to Einstein gravity (see Eq.~\eqref{eq-1112231}).\footnote{Recall that such higher curvature corrections will be induced in the semiclassical limit anyways, so it would be unnatural to ignore them, though we will show that they are inconsequential for the argument.} They only depend on the local geometry of $\eth\mathcal{W}$ (intrinsic and extrinsic geometry of $\eth \mathcal{W}$, and also the ambient $R_{ijkl}$ there) and the ambient local geometry. Lastly, the term $Q$ denotes all contributions of $o(\ell_S^0)$.

We can obtain an explicit expression for $\Theta_\lambda$ from Eq.~\eqref{eq-24tr114t} (see the definition in Eq.~\eqref{eq-11cx5}):
\begin{align}\label{eq-4n5895}
    \Theta_\lambda(y^a) = \theta_\lambda(y^a) +\cdots+\frac{4\ell_S^{d-2}}{\sqrt{h_\lambda}}\left.\frac{\delta S^{(1)}_{\rm ren}}{\delta V(y^a)}\right\rvert_\lambda + o(\ell_S^{d-2}),
\end{align}
where ellipsis denote contributions from $S_{\rm local}$, which are therefore suppressed at leading by a factor of $\ell_S^2 / R$. Therefore, the rQFC condition $\Theta_\lambda(y^a)=0$ implies:
\begin{align}\label{eq-6744}
    \theta_\lambda(y^a) = -\frac{4\ell_S^{d-2}}{\sqrt{h_\lambda}}\left.\frac{\delta S_{\rm ren}^{(1)}}{\delta V(y^a)}\right\rvert_\lambda + o(\ell_S^{d-2}),
\end{align}
because this equation in particular implies $R\sim \ell_S^{2-d}$, which when plugged into Eq.~\eqref{eq-4n5895} makes the ellipsis terms subleading to the renormalized entropy terms.

We can also compute $\partial_\lambda \Theta_\lambda$:
\begin{align}\label{eq-v31vv}
     \partial_\lambda \Theta_\lambda = - \frac{\theta_\lambda^2}{d-2} +\cdots - \frac{4\ell_S^{d-2} \theta_\lambda}{\sqrt{h_\lambda}} \left.\frac{\delta S_{\rm ren}^{(1)}}{\delta V(y^a)}\right\rvert_\lambda + \frac{4 \ell_S^{d-2}}{\sqrt{h}} \partial_\lambda \left.\frac{\delta S_{\rm ren}^{(1)}}{\delta V(y^a)}\right\rvert_\lambda +o(\ell_S^{d-2}),
\end{align}
where we have used the flatness of the ambient geometry in a neighborhood of the deformation to set the shear-squared $\varsigma^2 =0$, and $R_{ij}=0$. Furthermore, the ellipsis which come from $S_{\rm local}$ are suppressed at least by $\ell_S^2 / R^2$, so $R\sim \ell_S^{2-d}$ implies that the ellipsis terms are in particular $o(\ell_S^{2(d-2)})$.

The statement of the rQFC for the ball-shaped region is then:
\begin{align}\label{eq-2v413}
    \Theta_\lambda(y^a)=0 \implies \partial_\lambda \Theta_\lambda(y^a) &= \frac{16(d-3)}{d-2} \ell_S^{2(d-2)} \left(\left.\frac{1}{\sqrt{h_\lambda}}\frac{\delta S^{(1)}_{\rm ren}}{\delta V(y^a)}\right\rvert_\lambda\right)^2\nonumber\\
    &+ \frac{4 \ell_S^{d-2}}{\sqrt{h_\lambda}} \partial_\lambda \left.\frac{\delta S^{(1)}_{\rm ren}}{\delta V(y^a)} \right\rvert_\lambda + \partial_\lambda \left(\frac{4 \ell_S^{d-2}}{\sqrt{h_\lambda}}\frac{\delta Q}{\delta V(y^a)}\right) +o(\ell_S^{2(d-2)}) \leq 0.
\end{align}
The $o(\ell_S^{2(d-1)})$ terms in particular consist of the contributions from the corrections in Eq.~\eqref{eq-6744}, and also of Eq.~\eqref{eq-v31vv} (once we impose $R \sim \ell_S^{2-d}$ as a consequence of Eq.~\eqref{eq-6744}). In Appendix~\ref{sec:Q}, we argue that the contribution from $Q$ in Eq.~\eqref{eq-2v413} is also $o(\ell_S^{2(d-2)})$, that is:
\begin{align}\label{eq-assume}
    \partial_\lambda \left(\frac{4 \ell_S^{d-2}}{\sqrt{h_\lambda}}\frac{\delta Q}{\delta V(y^a)}\right) = o (\ell_S^{2(d-2)}),
\end{align}
which we assume from here on, and therefore we no longer explicitly write the $Q$ term in Eq.~\eqref{eq-2v413}.

We now discuss the von Neumann entropy derivative terms (the first two terms the RHS of Eq.~\eqref{eq-2v413}). We need only to evaluate them at leading order in $\ell_S$, though higher $\ell_S$ corrections can also be defined by taking into account the proper dressing of $O^{(q)}_f$, and analyzing the backreaction. Despite the unitary \eqref{eq-112131} not changing the geometry near the locus of null deformation, the $O^{(q)}_f$ in Eq.~\eqref{eq-1c1311} couples the inside and outside of $\mathcal{W}$, and therefore can change the von Neumann entropy derivatives non-trivially. In particular, in the regime $L \ll R$, we expect:
\begin{align}\label{eq-23c1}
    \frac{1}{\sqrt{h}}\frac{\delta S^{(1)}_{\rm ren}}{\delta V(y^a)} \sim \frac{\epsilon}{L^{d-1}}.
\end{align}
Let us justify the scaling in Eq.~\eqref{eq-23c1}. In the Minkowski vacuum, the only scale which could govern the LHS of Eq.~\eqref{eq-23c1} is $R$. After changing the state to $\ket{\Psi}$, we expect that the answer can now depend non-trivially on $\epsilon$, $L$, and $R$. Suppose now that we first take $R$ to be very large (as required by the condition $L \ll R$). Then, we expect that the dependence on $R$ will drop out of the entropy derivative. Furthermore, in this limit, we can approximate the ball-shaped region by a Rindler wedge. Therefore, at $O(\epsilon^0)$ the entropy derivative must be zero.\footnote{This is simply a result of symmetry in the Rindler wedge. The transverse integral of the functional derivative in Eq.~\eqref{eq-23c1} gives the entropy shape derivative associated with moving the Rindler wedge in the null direction to an identical Rindler wedge. Therefore, the total entropy derivative must be zero. By the transverse symmetry of the Rindler wedge (in the Minkowski vacuum), the functional derivative must vanish as well.} Therefore, the first non-trivial term appears at $O(\epsilon)$, and the powers of $L$ in Eq.~\eqref{eq-23c1} is enforced by dimensional analysis. Note that it was important in this analysis that the theory is a CFT, so the only scales we need to consider are that of the region and the state. Of course the scaling dimension of the operator $O^{(q)}_f$ matters as well, but those are dimensionless and will only affect the constant of proportionality in Eq.~\eqref{eq-23c1}.

\subsection*{Conformally transforming $\partial_\lambda(\delta S^{(1)}_{\rm ren}/\delta V_\lambda(y^a))$ into its Rindler wedge analogue}

We now discuss the term $\partial_\lambda(\delta S^{(1)}_{\rm ren}/\delta V_\lambda(y^a))$ in Eq.~\eqref{eq-2v413}, which is slightly more involved. In~\cite{Balakrishnan:2019gxl}, this quantity was computed explicitly in an $\epsilon$ expansion for null deformations of a Rindler wedge. We take advantage of a conformal transformation which maps $\mathcal{W}$ to the Rindler wedge, and its null deformations. Starting from Minkowski coordinates $X^i$ of Eq.~\eqref{eq-4t1v241}, we can consider a transformation to new coordinates $\tilde{X}^i$ using a special conformal transformation followed by a translation:
\begin{align}\label{eq-4211}
    X^i = \frac{\tilde X^i - \tilde X^2 C^i}{1-2 (C_i \tilde X^i) + C^2 \tilde X^2} + 2 R^2 C^i,
\end{align}
where 
\begin{align}\label{eq-24n111}
    C^i = \left(\tilde{X}^0= 0, \tilde{X}^1 =\frac{1}{2R} , 0 , \cdots\right),
\end{align}
and the inner products are with Minkowski metric $\eta_{ij}$, e.g., $C_i X^i = \eta_{ij}C^i X^j$.\footnote{In Eq.~\eqref{eq-24n111}, since $\chi=0$ we do not need to specify the other $d-2$ coordinates.} The relation between the metric in these coordinates is given by
\begin{align}\label{eq-3241}
    \Omega^{2}(X) \left[-dudv+R^2\left(1+\frac{v-u}{2R}\right)^2d\Omega_{d-2}^2\right] =  -d\tilde{u} d\tilde{v} + \sum_{a=1}^{d-2}d\tilde{y}_a^2,
\end{align}
where
\begin{align}
    \Omega(X) = 1-2 (C_i X^i) + C^2 X^2.
\end{align}

Note that in Eq.~\eqref{eq-3241}, the metric in the brackets on the LHS and RHS are both Minkowski metrics, but the coordinates are such that $u=v=0$ is a $(d-2)$ sphere of size $R$, while $\tilde{u}=\tilde{v}=0$ is an infinite plane parametrized by polar coordinates $\rho$ and the additional $(d-2)$ angles. In analogy with the $X^i$ coordinates, we refer to the transverse directions in coordinates $\tilde{X}^i$ collectively as $\tilde{y}^a$, though as before we single out a radial coordinate $\tilde{\rho}$ using:
\begin{align}\label{eq-131145f}
    \sum_{a=1}^{d-2}d\tilde{y}_a^2 = d \tilde{\rho}^2 + \tilde{\rho}^2 d \Omega_{d-3}^2.
\end{align}

The coordinate transformation \eqref{eq-4211} maps the $\mathcal{W}$ wedges, i.e., the domain of dependence of a ball-shaped region and its null deformations specified by $v=V(y^a)$, into $\tilde{\mathcal{W}}$, i.e., a Rindler wedge and its corresponding null deformations specified by $\tilde{v}=\tilde{V}(\tilde{y}^a)$ (see Fig.~\ref{fig:conformal_transform}). Since a Weyl transformation can remove the factor of $\Omega^2(X)$ in Eq.~\eqref{eq-3241}, this has implications for a CFT. In particular, it turns out that the renormalized entropy is a conformal invariant~\cite{Koeller:2015qmn, Graham:1999pm}.That is,
\begin{align}\label{eq-21v5g}
\tilde{S}^{(1)}_{\rm ren}(\tilde{V}(\tilde{y}^a))=S^{(1)}_{\rm ren}(V(y^a)),
\end{align}
where $\tilde{S}^{(1)}_{\rm ren} (\tilde{V}(\tilde{y}^a))$ is the renormalized entropy in the conformally transformed state of a certain deformation of a Rindler wedge, determined by $\tilde{V}(\tilde{y}^a)$. In even dimensions, there is in general an additive anomaly term in Eq.~\eqref{eq-21v5g}. This term will not change any of our arguments going forward, and we will absorb it in the definition of $\tilde{S}_{\rm ren}$.

We will restrict to $V(y)$ which only depends on $\chi$, where the map between $V(y^a)$ and $\tilde{V}(\tilde{y}^a)$ simplifies. Explicitly, it is easy to derive that:
\begin{align}\label{eq-1v256h}
    V(\chi) = \frac{\tilde{V}(\tilde{\rho})}{1+ \tilde{V}(\tilde{\rho})/2R} + O\left(\frac{\tilde{\rho}^2}{R^2}\right).
\end{align}
Now, suppose we define a natural $\lambda$-parametrized family of null-deformed Rindler wedges by:
\begin{align}
\tilde{V}_{\lambda}(\tilde{y}^a) = \lambda \exp \left(\frac{1}{1-\Sigma^2 / \tilde{\rho}^2}\right), ~~|\tilde{\rho}|\leq \Sigma,
\end{align}
and $\tilde{V}_{\lambda}(\tilde{y}^a) =0$ elsewhere. It then follows using the chain rule that:
\begin{align}\label{eq-23c12}
    \partial_\lambda\left.\frac{\delta \tilde{S}^{(1)}_{\rm ren}}{\delta \tilde{V}_\lambda(\tilde{y}^a)}\right\rvert_\lambda = \partial_{\lambda}\left.\frac{\delta S^{(1)}_{\rm ren}}{\delta V_\lambda(y^a)}\right\rvert_{\lambda} + \frac{1}{R} \left.\frac{\delta S^{(1)}_{\rm ren}}{\delta V_{\lambda}(y^a)}\right\rvert_{\lambda},
\end{align}
It is desirable to rewrite Eq.~\eqref{eq-23c12} using Eq.~\eqref{eq-2314v25v} as instead:
\begin{align}\label{eq-23c122}
    \partial_{\lambda}\left.\frac{\delta \tilde{S}^{(1)}_{\rm ren}}{\delta \tilde{V}(\tilde{y}^a)}\right\rvert_{\lambda} = \partial_{\lambda}\left.\frac{\delta S^{(1)}_{\rm ren}}{\delta V(y^a)}\right\rvert_\lambda + \frac{2}{d-2} \theta_\lambda \left.\frac{\delta S^{(1)}_{\rm ren}}{\delta V(y^a)}\right\rvert_\lambda.
\end{align}
Using this, we can re-write the second term in Eq.~\eqref{eq-2v413} of the other two terms in Eq.~\eqref{eq-23c122}. After simplification, the statement of rQFC can be written as:
\begin{align}\label{eq-1c15655}
    \Theta_\lambda(y^a)=0 \implies \partial_\lambda \Theta_\lambda(y^a) = \frac{16(d-1)}{d-2} &\ell_S^{2(d-2)} \left(\left.\frac{1}{\sqrt{h_\lambda}}\frac{\delta S^{(1)}_{\rm ren}}{\delta V(y^a)}\right\rvert_{\lambda}\right)^2\nonumber\\
    &+ \frac{4 \ell_S^{d-2}}{\sqrt{h_\lambda}} \partial_{\lambda} \left.\frac{\delta \tilde{S}^{(1)}_{\rm ren}}{\delta \tilde{V}(\tilde{y}^a)} \right\rvert_{\lambda} +o(\ell_S^{2(d-2)}) \leq 0.
\end{align}
Note that the first term on the RHS is positive, therefore the rQFC enforces that the second term must be sufficiently negative.\footnote{Note that the second term on the RHS of Eq.~\eqref{eq-1c15655} is a Rindler wedge quantity in a state where $\langle T_{ij} \rangle k^i k^j=0$ in a neighborhood of the null deformation. Therefore, the QNEC merely implies that this term is negative (see Eq.~\eqref{eq-131q5}). However, in this setting, we see that the rQFC demands more than it being negative. It must be negative and large enough in magnitude.} We discuss an explicit expression for this term next.

\subsection*{Explicit expression for $\partial_{\lambda} (\delta \tilde{S}^{(1)}_{\rm ren} / \delta \tilde{V}(\tilde{y}^a))\rvert_{\tilde{V}_{\lambda}}$}

In~\cite{Balakrishnan:2019gxl}, the LHS of Eq.~\eqref{eq-23c122} was calculated in $d>2$.\footnote{In fact, the expression which we cite is true for any QFT, not just CFTs.} We describe the setup, and present this expression from~\cite{Balakrishnan:2019gxl}, and relegate a sketch of its derivation to Appendix~\ref{sec:EE}.

Let $\tilde{O}^j_f$ be the result of the operator $O_{(q)}^j$ defined in Eqs.~\eqref{eq:O} and \eqref{eq-1c1311} after conformal transformation~\eqref{eq-4211}. We consider $\tilde{\ket{\Psi}}$ the near vacuum state in $d>2$ Minkowski spacetime resulting from the conformal transformation of $\ket{\Psi}$ defined in Eq.~\eqref{eq-112131}:
\begin{align}\label{eq-13122}
    \tilde{\ket{\Psi}} = \sum_{q=1}^{c} e^{i \epsilon \tilde{O}^{(q)}_f} \ket{0},
\end{align}
where $\tilde{O}_f = \sum_{j=1}^c \tilde{O}^j_f.$


A main result of~\cite{Balakrishnan:2019gxl} is that (see Appendix~\ref{sec:EE} for a review):
\begin{align}\label{eq-1313c4r4}
\frac{1}{\sqrt{\tilde{h}_\lambda}}\partial_{\lambda}\left.\frac{\delta \tilde{S}^{(1)}_{\rm ren}}{\delta \tilde{V}(\tilde{y}^a)}\right\rvert_{\lambda=0} = -\frac{\epsilon^2}{2} \int d^{d-2} \tilde{y}' \sqrt{\tilde{h}}~\partial_{\lambda} \tilde{V}_{\lambda}(\tilde{y}'^a)~\int_{-\infty}^{\infty} ds~e^s\langle \tilde{O}^{(1)}_{f} \mathcal{E}(\tilde{y}'^a) \mathcal{E}(\tilde{y})\tilde{O}^{(1)}_{f_s} \rangle + O(\epsilon^3),
\end{align}
where $\mathcal{E}$ denotes the averaged null energy operator placed at transverse position $\tilde{y}$:
\begin{align}
    \mathcal{E} (\tilde{y}^a) = \int_{-\infty}^\infty d\tilde{v}~T_{\tilde{v}\tilde{v}}(\tilde{u}=0,\tilde{v},\tilde{y}^a),
\end{align}
where $T_{\tilde{v}\tilde{v}}$ denotes the indicated component of the stress-energy tensor, and $\tilde{O}^{(1)}_{f_s}$ is a result of evolving $\tilde{O}^{(1)}_f$ by a global boost operator by an amount $s$:
\begin{align}
\tilde{O}^{(1)}_{f_s} = e^{-iKs} \tilde{O}^{(1)}_f e^{iKs}.
\end{align}
where $K$ is the global boost generator. This can be arranged by boosting the profile $f$ to its boosted version $f_s$.~\footnote{For $\lambda \neq 0$, $O^{(1)}_{f_s}$ needs to be replaced by $e^{iKs} \tilde{O}^{(1)}_f e^{-iKs}$ where $K$ is the full modular Hamiltonian associated to the cut $\lambda$. Our conclusions are easily generalizable to arbitrary $\lambda$ by this replacement.}

We are interested in deriving how the LHS of Eq.~\eqref{eq-1313c4r4} scales with the parameters involved. Specifically, let us discuss how the RHS term at $O(\epsilon^2)$ scales with $\Sigma$, the transverse width of the deformation profile, and the scale $L$ associated to the state (as specified by the size of the smearing function in Eq.~\eqref{eq-1313c4r4}). The dependence on $\Sigma$ is controlled by the behavior of the integrand as $\mathcal{E}$'s approach each other. In particular it is easy to verify that:
\begin{align}\label{eq-qwjrg}
    \frac{1}{\sqrt{\tilde{h}_\lambda}}\partial_{\lambda}\left.\frac{\delta \tilde{S}^{(1)}_{\rm ren}}{\delta \tilde{V}(\tilde{y}^a)}\right\rvert_{\lambda=0} = O\left(\frac{\epsilon^2 \Sigma^{d-2-\delta}}{L^{2d-2-\delta}}\right)
\end{align}
where:
\begin{align}\label{eq-b3rg}
    \int_{-\infty}^{\infty}ds~e^s\langle \tilde{O}^{(1)} \mathcal{E}(\tilde{y}'^a) \mathcal{E}(\tilde{y}^a)\tilde{O}^{(1)}_{f_s} \rangle \stackrel{\tilde{y}'\to \tilde{y}}=O(|\tilde{y}-\tilde{y}'|^{{-\delta}})
\end{align}
where $|\tilde{y}-\tilde{y}'|$ denotes proper distance between the indicated points. In Eq.~\eqref{eq-qwjrg}, the factors of $L$ are put in by dimensional analysis.


We have now determined how the relevant terms in Eq.~\eqref{eq-1c15655} scales with the parameters in our problem, and are in a position to extract implications of the rQFC.

\subsection*{Putting everything together: rQFC implies a new bound}

Let us summarize the order of magnitude of the relevant terms in Eq.~\eqref{eq-1c15655}:
\begin{align}
    \frac{4(d-1)}{d-2} \ell_S^{2(d-2)} \left(\left.\frac{1}{\sqrt{h_\lambda}}\frac{\delta S^{(1)}_{\rm ren}}{\delta V(y^a)}\right\rvert_{\lambda=0}\right)^2 &= O\left(\epsilon^2 \frac{\ell_S^{2(d-2)}}{L^{2(d-1)}}\right)\label{eq-ddd1}>0,\\
    \frac{4 \ell_S^{d-2}}{\sqrt{h_\lambda}} \partial_{\lambda} \left.\frac{\delta \tilde{S}^{(1)}_{\rm ren}}{\delta \tilde{V}(\tilde{y}^a)} \right\rvert_{\lambda=0} &= O\left(\epsilon^2 \frac{\ell_S^{d-2}\Sigma^{d-2-\delta}}{L^{2d-2-\delta}}\right)<0,\label{eq-ddd2}
\end{align}
where in the first line the generic condition:
\begin{align}\label{eq-sg41fef}
    \left.\frac{1}{\sqrt{h_\lambda}}\frac{\delta S^{(1)}_{\rm ren}}{\delta V(y^a)}\right\rvert_{\lambda=0} = \left.\frac{1}{\sqrt{h_\lambda}}\frac{\delta \tilde{S}^{(1)}_{\rm ren}}{\delta \tilde{V}(\tilde{y}^a)}\right\rvert_{\lambda=0} = O\left(\frac{\epsilon}{L^{d-1}}\right)
\end{align}
was used. The first equality in Eq.~\eqref{eq-sg41fef} simply comes from the conformal transformation. And the second the inequality is the result of the QNEC, which in this particular setting, follows directly from Eq.~\eqref{eq-1313c4r4}.

For the rQFC to be valid, Eq.~\eqref{eq-ddd1} must not exceed Eq.~\eqref{eq-ddd2} in magnitude, since the former is positive. Furthermore, for our analysis so far to be valid, we need to remain in the semiclassical regime which requires all length scales in the problem to be much larger than $\ell_S$. In particular, we must satisfy $\Sigma \gg \ell_S$, while also satisfying:
\begin{align}
    \left(\frac{\Sigma}{L}\right)^{d-2-\delta} \gtrsim \left(\frac{\ell_S}{L}\right)^{d-2}.
\end{align}
Therefore, the rQFC implies that $\delta\geq0$. We can summarize this bound as:
\begin{align}\label{eq-newnew}
        \left.\frac{1}{\sqrt{h_\lambda}}\frac{\delta \tilde{S}^{(1)}_{\rm ren}}{\delta \tilde{V}(\tilde{y}^a)}\right\rvert_{\lambda=0} = O\left(\frac{\epsilon}{L^{d-1}}\right) \implies \frac{1}{\sqrt{\tilde{h}_\lambda}}\partial_{\lambda}\left.\frac{\delta \tilde{S}^{(1)}_{\rm ren}}{\delta \tilde{V}(\tilde{y}^a)}\right\rvert_{\lambda=0} \leq O(\frac{\epsilon^2\Sigma^{d-2}}{L^{2d-2}}).
\end{align}
By the expression on the RHS, we mean that the second derivative of the von Neumann entropy cannot approach zero faster than $\Sigma^{d-2}$ as $\Sigma \to 0$.
This bound is stronger than the QNEC in this class of states, since the QNEC merely requires that the RHS of \eqref{eq-newnew} is non-positive.

Let us end by suggesting a more universal bound which we speculate might be true in all QFTs and implies the above-mentioned bounds. Roughly speaking, our bound \eqref{eq-newnew} comes from the rQFC enforcing that the QNEC does not get close to saturation as $\Sigma \to 0$ by a larger power than $\Sigma^{d-2}$, whenever $\delta \tilde{S}^{(1)}_{\rm ren}/\delta\tilde{V}(\tilde{y}^a)\neq 0$ at the relevant order in $\epsilon$. We have made two simplifying assumptions to extract this bound. We restricted to near vacuum-states, and, in particular, to ones in which $\langle T_{ij} \rangle=0$ in a neighborhood of the deformation by picking our smearing function $f$ to have no support there. Certainly, one can consider states with the latter property that are not near-vacuum, e.g., by setting $\epsilon = 1$ in unitaries \eqref{eq-112131}. We then still expect some verison of the above-mentioned bound on the magnitude of \eqref{eq-ddd2} to hold. The only difference is that we cannot as directly connect it to the product of $\mathcal{E}$ operators. Perhaps, the restriction to $\langle T_{ij} \rangle=0$ at the locus of null deformation is not essential for the argument either, and is only a convenient choice for the computation. We therefore speculate here that the following universal bound holds in the Rindler wedge for all QFTs:
\begin{align}\label{eq-SQNEC1}
    2\pi \langle \tilde{T}_{ij}\rangle \tilde{k}^i \tilde{k}^j -\frac{1}{\sqrt{\tilde{h}_\lambda}}\partial_\lambda \left.\frac{\delta \tilde{S}_{\rm ren}}{\delta \tilde{V}(\tilde{y}^a)}\right\rvert_\lambda \stackrel{\Sigma \to 0}\geq \kappa \Sigma^{d-2} \left(\frac{1}{\sqrt{\tilde{h}_\lambda}} \left.\frac{\delta \tilde{S}_{\rm ren}}{\delta \tilde{V}(\tilde{y}^a)}\right\rvert_\lambda\right)^2, ~~~~~~\text{(Speculative)}
\end{align}
where $\kappa>0$ needs to be determined. Here, tilde denotes that everything is evaluated on the Rindler wedge, and $k^i$ denotes the (surface-orthogonal) null vector field on $\partial^+\tilde{\mathcal{W}}$ which generates the $\tilde{V}_\lambda(y^a)$ flow.

When applied to our near-vacuum states \eqref{eq-112131} in a CFT, the bound \eqref{eq-SQNEC1} implies Eq.~\eqref{eq-newnew}. We leave further exploration of it to future work.

\section*{Acknowledgements}

We are grateful to Ido Ben-Dayan, Raphael Bousso, Tom Faulkner, Luca Iliesiu, Adam Levine, Raghu Mahajan, Juan Maldacena, Hervé Partouche, Geoff Penington, David Simmons-Duffin, Zhenbin Yang, Aron Wall, and Alexander Zhiboedov for insightful discussions. We especially thank David Simmons-Duffin and Alexander Zhiboedov for very helpful exchanges about lightray operators, Aron Wall for comments on an earlier draft, and Raghu Mahajan for comments on an earlier draft, and for pointing out an error in the first version of this paper.

V.F. would like to thank the Leinweber Institute for Theoretical Physics and Department of Physics for their hospitality during the early stages of this work. V.F. acknowledges financial support from the European Research Council (grant BHHQG-101040024). Views and opinions expressed are however those of the authors only and do not necessarily reflect those of the European Union or the European Research Council. Neither the European Union nor the granting authority can be held responsible for them.
P.T. was supported by the National Science Foundation Graduate Research Fellowship under Grant No. DGE-2146752. Any opinions, findings, and conclusions or recommendations expressed in this material are those of the authors and do not necessarily reflect the views of the National Science Foundation. ASM is supported by the LITP at UC Berkeley, Department of Energy through DE-SC0019380, and DE-FOA0002563, by AFOSR award FA9550-22-1-0098 and by a Sloan Fellowship. SK is supported by the LITP at UC Berkeley, Department of Energy through QuantISED Award DE-SC0019380 and Award DE-SC0025293.

\appendix

\section{JT gravity as dimensional reduction}
\label{sec:reduc}

In addition to its solvability, one of the main reasons why JT gravity has received a lot of interest is its relation to higher-dimensional gravity. Indeed, the action \eqref{eq:JT} can be obtained from a dimensional reduction of the $d$-dimensional Einstein-Hilbert action with cosmological constant $\hat \Lambda$:
\begin{equation}
\label{eq:JTaction}
    I_{\rm EH}=\frac{1}{16\pi \hat{G}}\int_{\hat{\mathcal{M}}}d^{d}X\sqrt{-\hat{g}}\left[\hat R-\hat{\Lambda}\right] + \frac{1}{8\pi \hat{G}}\int_{\partial\hat{\mathcal{M}}} d^{d}Y\sqrt{-\hat{h}}\hat{K},
\end{equation}
where $\hat G$ is Newton's constant on the original $d$-dimensional manifold $\hat{\mathcal{M}}$ with coordinates $\{X^{M},~M=0,...,d-1\}$, $\hat{g}_{MN}$ is the metric tensor on $\hat{\mathcal{M}}$ and $\hat{g}$ its determinant, $\hat{R}$ is the Ricci scalar. $\{Y^{M},~M=0,...,d-2\}$ are the coordinates on the boundary $\partial\hat{\mathcal{M}}$ with induced metric $\hat h_{MN}$, and $\hat{K}$ is the trace of its extrinsic curvature. The radius of curvature $L_{\rm (A)dS}$ is related to the higher-dimensional cosmological constant $\hat{\Lambda}$ by $\hat{\Lambda}=\pm (d-1)(d-2)/L_{\rm (A)dS}^2$.

The JT gravity action \eqref{eq:JT} can be obtained from the metric ansatz\footnote{See, for example, Appendix A of~\cite{Svesko:2022txo} for a derivation.}
\begin{equation}
\label{eq:sphred}
    d s^2 = \hat{g}_{MN}d X^Md X^N = g_{ij}(x) d x^{i}d x^{j} + L_{\rm (A)dS}^2\Phi^{2/(d-2)}(x)d\Omega_{d-2}^2.
\end{equation}
The metric $g_{ij}$ and coordinates $\{x^{i},~i=0,1\}$ describe a two-dimensional manifold $\mathcal{M}$, such that $\hat{\mathcal{M}} =\mathcal{M}\cross\mathbb{S}^{d-1}$ with the dilaton field $\Phi$ encoding the size of the $(d-2)$-dimensional compact space $\mathbb{S}^{d-2}$. The action then reduces to
\begin{equation}\label{eq:dim_red_action}
\begin{split}
I_{\rm EH}=&\,\frac{1}{16\pi G}\int_{\mathcal{M}}d^2 x\sqrt{-g}\left[\Phi R-\hat{\Lambda} \Phi+\frac{(d-3)(d-2)}{L_{\rm (A)dS}^2}\Phi^{\frac{d-4}{d-2}}+\frac{d-3}{d-2}\frac{(\nabla \Phi)^2}{\Phi}\right]\\
&+\frac{1}{8\pi G}\int_{\partial\mathcal{M}}d y \sqrt{-h}\Phi K,
\end{split}
\end{equation}
where $R$ is the two-dimensional Ricci scalar, $\nabla$ the covariant derivative compatible with the metric $g_{ij}$, and $K$ the trace of the extrinsic curvature of the boundary $\partial\mathcal{M}$ of $\mathcal{M}$. The two-dimensional Newton constant $G$ is given by 
\begin{equation}\label{eq:2d_Newton_cste}
\frac{1}{G} = \frac{S_{d-2}(L_{\rm (A)dS})}{\hat{G}},
\end{equation}
where $S_{d-2}(L_{\rm (A)dS})=2\pi^{(d-1)/2}~L_{\rm (A)dS}^{d-2}/\Gamma((d-1)/2)$ is the surface area of the $(d-2)$-sphere of radius $L_{\rm (A)dS}$.

When $d=3$, the action \eqref{eq:dim_red_action} simplifies to \eqref{eq:JT}, and $\phi_0=0$. It can be interpreted as the dimensional reduction of an empty (A)dS background. When spherically reducing a higher-dimensional metric with $d>3$, the kinetic term can be removed by a Weyl rescaling $g_{ij}\rightarrow \frac{1}{d-1}\Phi^{-\frac{d-3}{d-2}}g_{ij}$. One then expands around $\Phi=\phi_0=1$, where the potential vanishes, and which leads to the action of an extremal black hole \cite{Svesko:2022txo}. We find the JT gravity action \eqref{eq:JT}, with $\Phi=\phi_0+\phi$ and $\phi\ll\phi_0$.

From the point of view of the two-dimensional background, $\phi_0$ is a topological term in the action and does not change the equations of motion. However, the dimensional reduction confers an interpretation to the dilaton as the area associated with points in $\mathcal{M}$. Let the wedge $\mathcal{W}$ be such that $\eth\mathcal{W}=\cup_i P_i$ with $\{P_i\}$ a set of points in $\mathcal{M}$. It is associated with a $(d-1)$-dimensional spherically symmetric wedge $\hat{\mathcal{W}}$ in $\hat{\mathcal{M}}$ through dimensional reduction. Then,
\begin{equation} 
\label{eq:JTarea}
\frac{A(\eth\hat{\mathcal{W}})}{4\hat{G}} = \sum_{i}\frac{\Phi(P_i)}{4G}. 
\end{equation}
Thus, we impose
\begin{equation} 
\Phi\geq 0.
\end{equation}
This leads to the definition of an expansion parameter for lightrays in the two-dimensional background, 
\begin{equation}\label{eq:theta}
    \theta = \frac{1}{\Phi}\frac{d\Phi}{d\lambda} = \frac{\Phi'}{\Phi},
\end{equation}
where $'=\frac{d}{d\lambda}$ denotes differentiation with respect to an affine parameter $\lambda$ along $\partial^+\mathcal{W}$.

Taking the derivative of \eqref{eq:theta} with respect to the affine parameter $\lambda$ yields the two-dimensional Raychaudhuri equation:
\begin{equation}\label{eq:Raychaudhuri_2d}
    \theta' = \frac{\Phi''}{\Phi} - \theta^2.
\end{equation}
From the classical equation of motion \eqref{eq:contracted_eom}, one sees that $\Phi''=-8\pi G~k^{i}k^{j}T_{ij}$ for any null vector $k^{i}$, which is non-positive by the null energy condition. Since $\Phi$ is positive, one gets a two-dimensional version of the classical focusing theorem: $\theta' \leq 0$. Equivalently, Eq.~\eqref{eq:Raychaudhuri_2d} can be obtained from the dimensional reduction of the Raychaudhuri equation in higher dimensions, with the metric ansatz \eqref{eq:sphred}. Indeed, let us consider a codimension-$1$ null congruence of curves in $d$ dimensions. For $d>2$, the Raychaudhuri equation governing the evolution of its expansion scalar $\theta$ reads:
\begin{equation}\label{eq:Raychaudhuri_n+1}
\frac{d\theta}{d\lambda}=-\frac{1}{d-2}\theta^2-\hat\varsigma_{MN}\hat\varsigma^{MN}+\hat\omega_{MN}\hat\omega^{MN}-\hat R_{MN} k^{M}k^{N},
\end{equation}
where $\hat \varsigma_{MN}$ and $\hat \omega_{MN}$ are the shear and twist tensors respectively, $\hat R_{MN}$ is the $d$-dimensional Ricci tensor computed from the metric $\hat g_{MN}$, and $k^M$ a null vector orthogonal to the congruence. For an $SO(d+1)$-symmetric congruence, compatible with the symmetry of the metric ansatz \eqref{eq:sphred}, $k^M$ has no component along the $(d-2)$ compact dimensions, so that $\hat R_{MN}k^M k^N=\hat R_{ij}k^{i}k^{j}$. Using the ansatz \eqref{eq:sphred}, one finds that
\begin{equation}
\hat{R}_{ij} = R_{ij} - \frac{1}{\Phi}\nabla_{i}\nabla_{j}\Phi + \frac{d-3}{d-2}\frac{1}{\Phi^2}\nabla_{i}\Phi\nabla_{j}\Phi,
\end{equation}
where $R_{ij}$ is the two dimensional Ricci tensor computed from the metric $g_{ij}$. Using this relation, the dimensional reduction of the Raychaudhuri Eq. \eqref{eq:Raychaudhuri_n+1} gives (in absence of shear and twist):
\begin{equation}
    \frac{d\theta}{d\lambda}=\frac{\Phi''}{\Phi}-\theta^2-R_{ij}k^i k^j.
\end{equation}
Since in two dimensions the Ricci tensor is proportional to the metric tensor and $k^{i}$ is null, $R_{ij}k^{i}k^{j}=0$, which indeed gives back Eq.~\eqref{eq:Raychaudhuri_2d}. 

\section{Sketch of the derivation of Eq.\texorpdfstring{\boldmath ~\eqref{eq-1313c4r4}}{4.27}}
\label{sec:EE}

Here, we sketch the derivation of Eq.~\eqref{eq-1313c4r4} from~\cite{Balakrishnan:2019gxl}. For a more comprehensive discussion, we refer the reader to that paper. It is understood that information-theoretic aspects of QFT are best described using operator algebras associated to subregions, and density matrices associated to subregions are in general ill-defined in the continuum limit~\cite{Haag:1996hvx}. Nevertheless throughout this section we assign density matrices to QFT subregions as a shortcut to obtain our result (one can imagine that this is possible after an appropriate regularization scheme). The final expression is writable in a manner which is regularization independent, and can be phrased in algebraic QFT language.

In~\cite{Balakrishnan:2019gxl}, the shape deformation of the relative entropy was considered in a near-vacuum state of the type we discuss in Sec.~\ref{sect:CFT}. Let $\mathcal{W}$ denote a null-deformed Rindler wedges (as opposed to the main text, do not use tilde in this appendix to refer to the Rindler wedge since we are not worried about a mix up with the ball-shaped region). A wedge $\mathcal{W}$ is determined by a profile $V(y^a)$ in Minkowski spacetime coordinates:
\begin{align}
    ds^2 = -du dv + \sum_{a=1}^{d-2} (dy^a)^2,
\end{align}
where $u$ and $v$ are null coordinates and $y^a$ denotes the transverse direction. We can also consider a $1$-parameter family of wedges $\mathcal{W}_\lambda$ with $\partial_\lambda V_\lambda \geq 0$.

Let us now denote by $\rho_\mathcal{W}$ and $\sigma_\mathcal{W}$ the density matrices associated to $\mathcal{W}$ in a near-vacuum and the vacuum states respectively. An explicit near-vacuum density matrix $\rho_\mathcal{W}$ can be constructed as follows:
\begin{align}
    \rho_\mathcal{W} = \sigma_\mathcal{W} + \delta \rho_\mathcal{W},
\end{align}
where
\begin{align}\label{eq-1954t}
    \delta \rho_{\mathcal{W}} = \epsilon(\sigma_\mathcal{W} O_f + O_f \sigma_\mathcal{W}).
\end{align}
Here, $\epsilon\ll 1$, and $O_f$ is some smeared local Hermitian operator.

It is understood that the quantity appearing in the QNEC can be written in terms of shape deformation of the relative entropy in the following way~\cite{Bousso:2015wca, Koeller:2017njr, Casini:2017roe}:
\begin{align}\label{eq-116bbf}
    \partial_\lambda \left.\frac{\delta S_{\rm rel}(\rho_\mathcal{W}|\sigma_\mathcal{W})}{\delta V(y^a)}\right\rvert_\lambda = 2\pi \langle T_{vv} \rangle - \partial_\lambda \left.\frac{\delta S_{\rm ren}(\rho_\mathcal{W})}{\delta V(y^a)}\right\rvert_\lambda,
\end{align}
where $\langle T_{vv} \rangle$ denotes expectation value in the near-vacuum state $\rho$. We will compute the LHS of Eq.~\eqref{eq-116bbf}. For nearby density matrices, one has the following expansion (see e.g., Appendix B of~\cite{Faulkner:2017tkh}):
\begin{align}
    S_{\rm rel}(\rho_\mathcal{W}|\sigma_\mathcal{W}) = -\int_{-\infty}^{\infty} ds \frac{1}{8\sinh^2(\frac{s+i\epsilon}{2})} \Tr[ \sigma_\mathcal{W}^{-1} \delta\rho_\mathcal{W} \sigma_\mathcal{W}^{\frac{is}{2\pi}}\delta\rho_\mathcal{W} \sigma_\mathcal{W}^{-\frac{is}{2\pi}}] + O(\delta \rho^3).
\end{align}
Plugging in Eq.~\eqref{eq-1954t} and doing simple manipulations one arrives at:
\begin{align}\label{eq-41v6h}
    S_{\rm rel}(\rho_\mathcal{W}|\sigma_\mathcal{W}) = -\epsilon^2 \int_{-\infty}^{\infty} ds \frac{1}{8\sinh^2(\frac{s+i\epsilon}{2})} \langle O_f e^{i K s} O_f \rangle + O(\epsilon^3),
\end{align}
where $2\pi K_\mathcal{W} = \log \sigma_\mathcal{W} - \log \sigma_{\mathcal{W}^{\text{complement}}}$, where $\mathcal{W}^{\text{complement}}$ denotes the causal complement of the wedge $\mathcal{W}$, denotes the full modular Hamiltonian of the vacuum state associated with $\mathcal{W}$. The expectation value in Eq.~\eqref{eq-41v6h} is computed in the vacuum state. The full modular Hamiltonian can be explicitly written as the following boost-like expression:
\begin{align}
    K_{\mathcal{W}} = 2\pi \int_{-\infty}^{\infty} dv \int d^{d-2}y^a~(v-V(y^a)) T_{vv}(u=0, v,y^a).
\end{align}
It turns how that modular Hamiltonians of different cuts of the Rindler wedge are related like below:
\begin{align}
    e^{-i K_\mathcal{W} s} e^{-i K_0 s} = e^{i (e^s-1) \int dy^a \int dv~V(y^a) T_{vv}(u=0,v,y^a)},
\end{align}
where $K_0$ denotes the full modular Hamiltonian for the original Rindler wedge, i.e., one corresponding to $V=0$.
This relations allows us to compute the shape derivative in Eq.~\eqref{eq-116bbf} explicitly from Eq.~\eqref{eq-41v6h}. Each derivative drops down factors of the averaged null energy operator $\mathcal{E}(y^a) = \int_{-\infty}^{\infty} dv~ T_{vv}(u=0,v,y^a) $ from the full modular Hamiltonian term, and we end up with:
\begin{align}
 \partial_\lambda \left.\frac{\delta S_{\rm rel}(\rho_\mathcal{W}|\sigma_\mathcal{W})}{\delta V(y^a)}\right\rvert_\lambda = \frac{\epsilon^2}{2}\int d^{d-2}y'^a \int ds ~\partial_\lambda V_\lambda(y'^a)  \langle O_f \mathcal{E}(y^a) \mathcal{E}(y'^a) e^{i K_{\mathcal{W}_\lambda} s} O_f \rangle + O(\epsilon^3).
\end{align}
Note that the QNEC implies that the RHS is positive. Also, since the full modular Hamiltonian of any $\mathcal{W}$ annihilates the vacuum, we can write the above as:
\begin{align}\label{eq-3v141rv}
 \partial_\lambda \left.\frac{\delta S_{\rm rel}(\rho_\mathcal{W}|\sigma_\mathcal{W})}{\delta V(y^a)}\right\rvert_\lambda = \frac{\epsilon^2}{2}\int d^{d-2}y'^a \int ds ~\partial_\lambda V_\lambda(y'^a)  \langle O_f \mathcal{E}(y^a) \mathcal{E}(y'^a) O_{f_s} \rangle + O(\epsilon^3),
\end{align}
where
\begin{align}
    O_{f_s} = e^{i K_{\mathcal{W}_\lambda} s} O_f e^{-i K_{\mathcal{W}_\lambda} s}
\end{align}
is a modular-evolved $O_f$ (simply a boost for $\mathcal{W}$ that are Rindler wedges).

Along with the condition $\langle T_{vv} \rangle_\rho=0$ which we engineered in Sec.~\ref{sect:CFT}, Eq.~\eqref{eq-3v141rv} gives us Eq.~\eqref{eq-1313c4r4}.

\section{The generalized entropy perturbative expansion in the semiclassical regime, and the \texorpdfstring{\boldmath $Q$}{Q} term}
\label{sec:Q}

As discussed in the introduction, Sec.~\ref{sect:semi}, and Sec.~\ref{sect:CFT}, in the perturbatively semiclassical gravity regime we expect the following expansion for the generalized entropy:
    \begin{align}\label{eq-124r1j}
    S_{\rm gen} (\mathcal{W}) = c \left(\frac{A(\eth\mathcal{W})}{4 \ell_S^{d-2}} + \cdots + S^{(1)}_{\rm ren} + Q(\mathcal{W}) \right),
\end{align}
where $Q(\mathcal{W})$ denotes corrections that are $o(\ell_S^0)$ in the expansion. In this appendix, we study $S_{\rm gen}$ in an explicit theory in which we derive this ansatz for a simple example, including the lesser-studied $Q$ term in Eq.~\eqref{eq-124r1j}. We then present an argument for the order of magnitude of the $Q$ term in the setup of Sec.~\ref{sect:CFT} (under Eq.~\eqref{eq-2v413}).

The particular class of theories in which we investigate this are $d$-dimensional holographic brane-world theories (see~\cite{Randall:1999ee,Randall:1999vf,Shiromizu:1999wj,Gubser:1999vj,Verlinde:1999xm,Karch:2000ct,Emparan:2006ni,Myers:2013lva}). In fact, these are theories in which the proof of the rQFC was given in~\cite{Shahbazi-Moghaddam:2022hbw}. The theory has a perturbative semiclassical limit in which a holographic CFT is weakly coupled to gravity (in fact, the gravity is induced by the CFT). The (Euclidean) action of this theory can be computed holographically in terms of that of a AdS$_{d+1}$ spacetime with an end-of-the-world brane with a specific tension (on which the brane-world theory lives). Explicitly:
\begin{align}\label{eq-234r1442}
    \log Z_{\rm brane-world} [g_{ij}] &= \frac{1}{16 \pi G_{d+1}}\int_{\rm bulk} d^dx dz~\sqrt{-G} \left(G^{\mu\nu}R_{\mu\nu} + \frac{d(d-1)}{\ell_{\rm AdS}^2}\right) \nonumber\\
    &+ \frac{1}{8 \pi G_{d+1}} \int_{\rm brane} d^dx~ \sqrt{-g} (K - T),
\end{align}
where $G_{\mu\nu}$ and $R_{\mu\nu}$ are the bulk metric and Ricci tensor respectively, $x$ denotes coordinates on the brane, and $z$ denotes the emergent bulk direction. Furthermore, $K$ is the extrinsic curvature of the brane, and $T \propto 1/\ell_{\rm AdS}$ is the brane tension. The boundary conditions administered on the brane is:
\begin{align}\label{eq-21215}
    K_{ij} - K g_{ij} = T g_{ij}.
\end{align}
Our perturbative semiclassical theory lives on the brane with metric $g_{ij}$, and is dual to classical Einstein gravity in the bulk (i.e., $G_{d+1} \to 0$ limit).\footnote{This setup is very similar to the way one defines the partition function of a holographic CFT with a cutoff in terms of AdS gravity with a ``Dirichlet brane''. The main difference here is that instead of the usual Dirichlet boundary conditions, we administer the boundary condition \eqref{eq-21215} in the presence of an end-of-the-world brane. The fictitious cutoff surface is then replaced by a brane whose location is determined dynamically. This ends up having the effect of turning on gravity on the brane.} To see that this indeed is a perturbative semiclassical theory, one can expand the action \eqref{eq-234r1442} in the limit of where the length scales on $g_{ij}$ are much larger than $\ell_{\rm AdS}$. The result is (see e.g.~\cite{Myers:2013lva}):
\begin{align}\label{eq-1c414}
    \log Z_{\rm brane-world} [g_{ij}] = \frac{1}{16 \pi G_d} \int_{\rm brane} d^d x \sqrt{-g} &\left(g^{ij} R_{ij} + \ell_{\rm AdS}^2 (\text{curvature-squared terms}) + \cdots\right)\nonumber\\
    &+ \log Z_{\rm CFT}[g_{ij}] + o(c \ell_{\rm AdS}^0),
\end{align}
where $R_{ij}$ is the intrinsic Ricci tensor on the brane, and
\begin{align}\label{eq-42113}
    \frac{1}{G_d} = \frac{\ell_{\rm AdS}}{(d-2)G_{d+1}}.
\end{align}
In Eq.~\eqref{eq-1c414}, the ellipsis in the parenthesis denotes further higher curvature corrections suppressed by appropriate powers of $\ell_{\rm AdS}$. We see that $\ell_{\rm AdS}$ acts as species scale in Eq.~\eqref{eq-1c414}. In fact, holographic CFTs contain large number of matter fields. For example, in $\mathcal{N}=4$ super Yang-Mills, we have:
\begin{align}\label{eq-3d13}
\ell_{\rm AdS}^{3} \propto N^2 G_{5}.
\end{align}
Combining Eqs.~\eqref{eq-3d13} and \eqref{eq-42113}, we obtain $\ell_{\rm AdS}^{2} = c G_4$, where $c \propto N^2$. More generally, we expect:
\begin{align}
    \ell_{\rm AdS}^{d-2} = c G_d,
\end{align}
where $c$ is proportional to the effective number of matter fields in the holographic CFT. We conclude that the bulk AdS scale $\ell_{\rm AdS}$ act as the brane-world theory's species scale.

\begin{figure}
    \centering
    \includegraphics[width=0.7\linewidth]{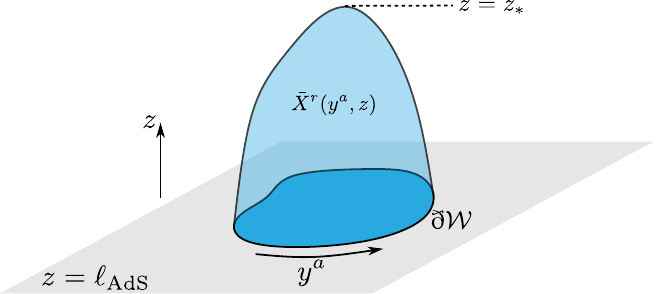}
    \caption{We can consider a more general region on the brane (at $z=\ell_{\rm AdS}$), whose $S_{\rm gen}$ is computed using the Bekenstein-Hawking entropy of a bulk extremal surface (denoted by embedding $\bar{X}^r(y^a,z)$) anchored to $\eth \mathcal{W}$ on the brane. The expansion of the reach of the extremal surface $z_*$ in the small $\ell_{\rm AdS}$ limit is significant for the subleading $Q$ terms in the $S_{\rm gen}$ expansion (see Eq.~\eqref{eq-0rhje}).}
    \label{fig:brane3d}
\end{figure}

In this theory, the generalized entropy can be defined in a way analogous to the holographic entanglement entropy prescription for CFTs~\cite{Emparan:2006ni,Myers:2013lva}. That is, given a wedge on the brane, we can define:
\begin{align}\label{eq-42v144}
    S_{\rm gen}(\mathcal{W}) = \frac{A(X)}{4 G_{d+1}},
\end{align}
where $A(X)$ is the area of the minimal bulk extremal surface which anchors to $\eth \mathcal{W}$ on the brane (see Fig.~\ref{fig:brane3d}). Similarly to action, we can expand this in the regime of small $\ell_{\rm AdS}$ in terms of brane quantities, and obtain an expansion:
\begin{align}\label{eq-0rhje}
    S_{\rm gen} (\mathcal{W}) = c \left(\frac{A(\eth\mathcal{W})}{4 \ell_{\rm AdS}^{d-2}} + \cdots + S^{(1)}_{\rm ren} + Q(\mathcal{W}) \right).
\end{align}
Here, the ellipsis denotes higher curvature Dong entropy terms which depend on the geometry of $\eth \mathcal{W}$, and $Q$ denotes corrections that are $o (\ell_{\rm AdS})$.

Let us derive this $S_{\rm gen}$ expansion in a specific example. Consider pure AdS dual to the brane-world in Minkowkski spacetime $\eta_{ij}$ and in the vacuum state. The bulk metric in standard Fefferman-Graham coordinates is:
\begin{align}
    ds^2 = \frac{\ell_{\rm AdS}^2}{z^2} \left(dz^2 + \eta_{ij} dx^i dx^j\right), ~~~~~z\geq \ell_{\rm AdS}.
\end{align}
This spacetime terminates at $z = \ell_{\rm AdS}$ where the end-of-the-world brane sits.

\begin{figure}
    \centering
    \includegraphics[width=0.5\linewidth]{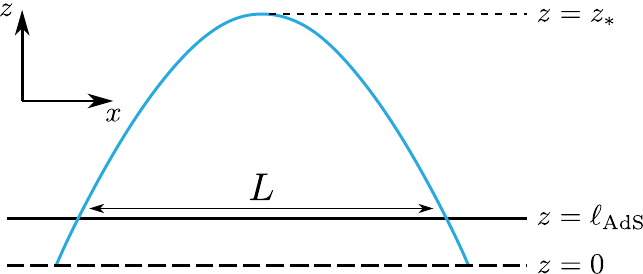}
    \caption{A simple setting in the holographic braneworld scenario is shown. The bulk is pure AdS, and the brane which sits at $z=\ell_{\rm AdS}$ (in Fefferman-Graham $z$), has Minkowski intrinsic metric. On the brane, we pick a belt region ($-L/2<x<L/2$) and compute its $S_{\rm gen}$ using the Bekenstein-Hawking entropy of a bulk extremal surface (shown in blue, and extending to $z_*$) which anchors on the edge of the belt region on the brane. We can then expand this $S_{\rm gen}$ in the small $\ell_{\rm AdS}$ limit (which is the theory's species scale), and recover an all-orders expansion.}
    \label{fig:brane2d}
\end{figure}

Let $(t,x,y^a)$ be standard Minkowski coordinates on the brane:
\begin{align}
    ds^2 = -dt^2 + dx^2 + \sum_{a=1}^{d-2} (dy^a)^2.
\end{align}
Now, consider a ``belt region'' on the brane specified by:
\begin{align}
    \{t=0, -\frac{L}{2}\leq x\leq \frac{L}{2} \}.
\end{align}
We furthermore assume that the $y^a$ extent of the slab $R$ is much larger than $L$, so is approximated as infinite for the purposes of the computation (see Fig.~\ref{fig:brane2d}).

We would like to compute the corresponding generalized entropy (for the wedge which is the domain of dependence of this region) using the prescription~\eqref{eq-42v144}. First, we must find the bulk extremal surface anchored to the slab. This can be specified by the embedding $z(x)$ which satisfies:
\begin{align}
    \frac{dz}{dx}
= \frac{\sqrt{z_*^{2(d-1)}-z^{2(d-1)}}}{z^d},
\end{align}
where $z_*$ is a constant and it marks the maximum value of $z$ that the bulk extremal surface reaches. It can be determined by solving:
\begin{align}\label{eq-857234}
    \frac{L}{2} = \int_{\ell_{\rm AdS}}^{z_*} dz \frac{z^{d-1}}{\sqrt{z_*^{2(d-1)} - z^{2(d-1)}}}.
\end{align}
We can compute the area of the extremal surface:
\begin{align}\label{eq-24r1v}
    S_{\rm gen} = \frac{\ell_{\rm AdS}^{d-1} R^{d-2}}{4 G_{d+1}}\!\int_{-L/2}^{\,L/2}\! dx\;
\frac{\sqrt{1+\left(\frac{dz}{dx}\right)^{2}}}{z^{d-1}} = \frac{1}{4 G_{d+1}}\left[\frac{2 \ell_{\rm AdS}^{d-1}}{d-2}\left(\frac{R}{\ell_{\rm AdS}}\right)^{d-2} + \frac{\sqrt{\pi}\Gamma(\frac{2-d}{2(d-1)})}{(d-1)\Gamma(\frac{1}{2(d-1)})} \ell_{AdS}^{d-1}\left(\frac{R}{z_*}\right)^{d-2}\right].
\end{align}

From Eq.\eqref{eq-857234}, one can solve for $z_*$ perturbatively in $\ell_S$. We find:
\begin{align}
    z_* = L \left[a_1 + a_2 \left(\frac{\ell_{\rm AdS}}{L}\right)^d+a_3 \left(\frac{\ell_{\rm AdS}}{L}\right)^{2d} + a_4 \left(\frac{\ell_{\rm AdS}}{L}\right)^{3d-2}+a_5 \left(\frac{\ell_{\rm AdS}}{L}\right)^{3d} + a_6 \left(\frac{\ell_{\rm AdS}}{L}\right)^{4d-2} + \cdots    \right],
\end{align}
where $a_j$ are certain (derivable) $d$-dependent numerical coefficients. We can then plug this into Eq.~\eqref{eq-24r1v} to obtain the desired answer:

\begin{equation}\label{eq-31c1}
\begin{split}
S_{\rm gen} =& c\left[\frac{2R^{d-2}}{4 \ell_{\rm AdS}^{d-2}} + b_1 \left(\frac{R}{L}\right)^{d-2} + b_2 \left(\frac{\ell_{\rm AdS}}{L}\right)^d  \left(\frac{R}{L}\right)^{d-2} + b_3 \left(\frac{\ell_{\rm AdS}}{L}\right)^{2d}  \left(\frac{R}{L}\right)^{d-2}\right.\\
&\qquad\left.+ b_4\left(\frac{\ell_{\rm AdS}}{L}\right)^{3d}  \left(\frac{R}{L}\right)^{d-2}+b_5 \left(\frac{\ell_{\rm AdS}}{L}\right)^{4d-2} \left(\frac{R}{L}\right)^{d-2} + \cdots \right],
\end{split}
\end{equation}
where $b_j$ are some $d$-dependent numerical coefficients. The first term in Eq.~\eqref{eq-31c1} is the usual area term ($A(\eth \mathcal{W})= 2 R^{d-2}$), and the second term is the renormalized von Neumann entropy. Note that there are no subleading $S_{\rm local}$ contributions between the two because the local geometry of $\eth \mathcal{W}$ is completely flat. Nevertheless, we see that there are non-trivial contributions at $o(\ell_S)$. These are all part of the contributions which we have called $Q$. We furthermore see that the contribution of $Q$ starts at $O(\ell_{\rm AdS}^d)$, consistent with the claim in Sec.~\ref{sect:CFT} that the contribution of $Q$ to $\partial_\lambda \Theta_\lambda$ is $o(\ell_S^{2(d-2)})$.

The above is highly non-trivial evidence in support of the claim on the magnitude of $Q$ contribution to $\partial_\lambda \Theta_\lambda$ (assumption in Eq.~\eqref{eq-assume}). But the setting of the belt example is not equivalent to the setup of Sec.~\ref{sect:CFT}. Below, we sketch an argument in support of the claim in a more general setting. For a more general region and state on the brane, the computation of $S_{\rm gen}$ would involve solving for an extremal surface embedding surface $(\bar{X}^r(y,z), y^a)$ for $r=t,x$ in the spacetime (see Fig.~\ref{fig:brane3d}):
\begin{align}
    ds^2 = \frac{\ell_{\rm AdS}^2}{z^2} \left(dz^2 + g_{ij}(z,x) dx^i dx^j \right), z>\ell_{\rm AdS},
\end{align}
where
\begin{align}\label{eq-35tb2}
    g_{ij}(z,x) =  \eta_{ij} + z^d g^{(d)}_{ij} (x) + o(z^d).
\end{align}
Eq.~\eqref{eq-35tb2} comes from solving the Einstein field equations perturbatively in $z$ near $z=0$. To find the brane metric, for example, we can evaluate $g_{ij}(z=\ell_{\rm AdS},x)$. Similarly to deriving the metric expansion~\eqref{eq-35tb2}, the near $z=0$ expansion for $\bar{X}^\mu$ has been analyzed in~\cite{Koeller:2015qmn,Akers:2017ttv}. The result is:
\begin{align}
    \bar{X}^r(y^a,z) = (X^{(0)})^r(y^a) + z^2 (X^{(2)})^r(y^a) + \cdots+ z^d \frac{\delta S_{\rm ren}}{\delta X^r (y^a)},
\end{align}
where $(X^{(2)})^r(y^a)$ depends only on the local geometry of $\eth \mathcal{W}$ on the brane. In the setup of Sec.~\ref{sect:CFT}, we can write this schematically as:
\begin{align}
    \bar{X}^r(y^a,z) = (X^{(0)})^r(y^a) + O\left(\frac{z^2}{R}\right) + O\left(\epsilon \frac{z^d}{L^{d-1}}\right).
\end{align}
Similarly, we can derive the following ansatz for $z_*$, the turn around point of the extremal surface in the $z$ direction\footnote{Here we are assuming that the Fefferman-Graham expansion is valid in a neighborhood which includes the turn around point of $\bar{X}$.}
\begin{align}\label{eq-6yh3}
    z_* = (z_*^{(0)}) + O\left(\frac{\ell_{\rm AdS}^2}{R}\right) + O\left(\frac{\epsilon \ell_{\rm AdS}^d}{L^{d-1}}\right).
\end{align}
We then need to evaluate:
\begin{align}
    S_{\rm gen} = \frac{1}{G_{d+1}} \int_{\ell_{\rm AdS}}^{z_*} dz \int d^{d-2} y ~ \sqrt{\bar{H}(y^a,z)},
\end{align}
where $\bar{H}(y^a,z)$ is the determinant of the intrinsic metric on the extremal surface. This can also be expaneded as follow:
\begin{align}\label{eq-dgh89}
\sqrt{\bar{H}(z,y^a)} = \frac{\ell_S^{d-1}}{z^{d-1}} \left[f^{(0)}(z,y) + \frac{z^2}{R} f^{(2)}(y,z) + \frac{\epsilon z^d}{L^{d-1}} f^{(d)}(y,z)+\cdots \right].
\end{align}
We need to explicitly compute:
\begin{align}\label{eq-2167uj}
    \frac{1}{c}S_{\rm gen} \propto \int d^{d-2} y \int_{\ell_{\rm AdS}}^{z_*} \frac{dz}{z^{d-1}} \left(f^{(0)}(z,y) + \frac{z^2}{R} f^{(2)}(y,z) + \frac{\epsilon z^d}{L^{d-1}} f^{(d)}(y,z)+\cdots\right),
\end{align}
where for the prefactors we used the above-mentioned relation between $G_{d+1}$, $G_{d}$, and $\ell_{\rm AdS}$, and $c$.

We can analyze the perturbative expansion in $\ell_{\rm AdS}$ of the integral~\eqref{eq-2167uj}. The terms in the integrand which blow up at $z=0$, can be analyzed locally near the lower bound $z=\ell_{\rm AdS}$ of the integral. For example, the first two terms give powers of $\ell_{\rm AdS}^{2-d}$ and $\ell_{\rm AdS}^{4-d}$ and will clearly have the form of a local integral in $y^a$. These constitute the $S_{\rm local}$ contributions in Eq.~\eqref{eq-0rhje}. To understand the (non-trivial) terms in $Q$, we need to understand the perturbative contributions related to the upper bound $z_*$. The leading order contribution from Eq.~\eqref{eq-6yh3} gives an $O(1)$ contribution inside of the integral. This constitute the $S_{\rm ren}$ term in Eq.~\eqref{eq-0rhje}. Further corrections would be then suppressed by the same factors that corrections to $z_*$ are suppressed by in Eq.~\eqref{eq-6yh3}, namely $O(\ell_{\rm AdS}^2 / R)$ and $O(\epsilon \ell_{\rm AdS}^d / L^{d-1})$ Both of these contributions in the context of Sec.~\ref{sect:CFT} where $R \sim \ell_{\rm AdS}^{2-d}$ become $O(\ell_{\rm AdS}^d)$.  Furthermore, since $\partial_\lambda \Theta$ has an additional multiplicative factor of $G_d$, the conclusion is that the contribution of the $Q$ term satisfies $O(\ell_{\rm AdS}^{2(d-1)})$ at most and is therefore compatible with the assumption $o(\ell_{\rm AdS}^{2(d-2)})$ (see Eq.~\eqref{eq-assume}).

\section{The $\mathcal{E} \times \mathcal{E}$ OPE}\label{sec:v1}

As discussed in Sec.~\ref{sect:CFT}, the rQFC implies that $\delta\geq0$ in the following expression:
\begin{align}\label{eq-b3rg11}
    \int_{-\infty}^{\infty}ds~e^2\langle \tilde{O}^{(1)}_{f} \mathcal{E}(\tilde{y}'^a) \mathcal{E}(\tilde{y}^a)\tilde{O}^{(1)}_{f_s} \rangle \stackrel{\tilde{y}'\to \tilde{y}}=O(|\tilde{y}-\tilde{y}'|^{{-\delta}})
\end{align}
The coincident limit $\tilde{y}' \to \tilde{y}$ is in particular controlled by the OPE of two $\mathcal{E}$'s. This is a less conventional OPE which involves light-ray operators (i.e., operators with support on null generators of the $\tilde{u}=0$ plane). See~\cite{Kologlu:2019mfz, Chang:2020qpj} for a proper discussion of light-ray operators and their OPEs.

\begin{figure}
    \centering
    \includegraphics[width=\linewidth]{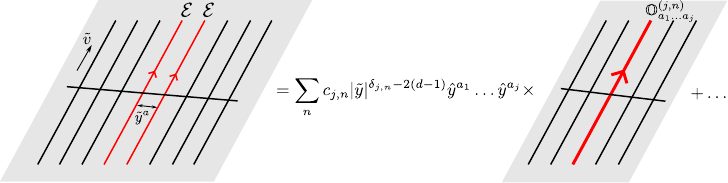}
    \caption{The OPE of two averaged null energy operators $\mathcal{E}(\tilde{y}^a) = \int_{-\infty}^{\infty} d\lambda~T_{\tilde{v}\tilde{v}}(\tilde{v}=\lambda, \tilde{y}^a)$ is shown. The RHS of the OPE consists of so-called light-ray operators $\mathbb{O}^{(j,n)}_{a_1 \cdots a_j}$, labeled by a transverse spin $j$, and corresponding to a certain $n$-th (even spin) Regge trajectory. In our states, the second derivative of the von Neumann entropy under the null deformations of the Rindler wedge are controlled by this OPE: the more singular (as $|\tilde{y}|\to 0$) the OPE, the smaller the second derivative becomes in the limit of deformation width going to zero.}
    \label{fig:lightOPE}
\end{figure}

Here, we will simply write down the relevant OPE and explain it in some detail. For simplicity let us set of the transverse points to zero. Let $|\tilde{y}| = \sqrt{\tilde{y}^i \tilde{y}_i}$, and $\hat{y}^a = \tilde{y}^a/|\tilde{y}|$. The OPE takes the form (see~\cite{Kravchuk:2018htv,Kologlu:2019mfz,Chang:2020qpj}):
\begin{align}\label{eq-234c1}
    \mathcal{E} (0) \mathcal{E}(\tilde{y}^a) \stackrel{|\tilde{y}| \to 0}{=}& \sum_{j,n} c_{j,n}~|\tilde{y}|^{\delta_{j,n} - 2(d-1)} \hat{y}^{a_1} \cdots \hat{y}^{a_j}~\mathbb{O}^{(j,n)}_{a_1 \cdots a_j} (0) + \text{descendents of $\mathbb{O}^{(j,n)}_{a_1 \cdots a_j}$},
\end{align}
where $\mathbb{O}^{(j,n)}_{a_1 \cdots a_j}$ denotes a so-called light-ray operator with scaling dimension $\delta_{j,n}$, and transverse spin $j$ (see Fig.~\ref{fig:lightOPE}). Each operator $\mathbb{O}^{(j,n)}_{a_1 \cdots a_j}$ is a non-local operator, and could be defined through smearing a pair local CFT operators in a certain way along a null geodesic. We will not cover the details of the light-ray OPE story here, but its structure \eqref{eq-234c1} is sufficient for us to extract a certain bound on the spectrum $\delta_{j,n}$. Conveniently, the spectrum $\delta_{j,n}$ could be described simply in terms of the spectrum of local CFT data, allowing us the ability to phrase the bound in terms of data of the local CFT spectrum. Let $J$ denote total spin (as opposed to transverse spin $j$), and let $\Delta^{\text{even}}_{j,n} (J)$ be the analytically continued in $J$ of the scaling dimension of the n-th Regge trajectory in the transverse spin $j$ sector. We then have~\cite{Kravchuk:2018htv,Kologlu:2019mfz, Chang:2020qpj}:
\begin{align}
    \delta_{j,n} =& \Delta^{\text{even}}_{j,n} (J=3)-1, ~~~ j=0,2,4.\label{eq-3jyf1}\\
    \delta_{j,n} =& \Delta^{\text{even}}_{j=4,n}(J=3+2p)-1,~~~j=4+2p,~(p\in \mathbb{Z}^+)\label{eq-3jyf2}.
\end{align}

It is easy to see from the OPE~\eqref{eq-234c1}, that the leading $\Sigma \to 0$ behavior of Eq.~\eqref{eq-1313c4r4} is given by the smallest $\delta_{j,n}$ which is contributing in the state. A particularly drastic way for the rQFC implication that $\delta\geq 0$ in Eq.~\eqref{eq-b3rg11} to be violated is if the quantity $\delta_0$ defined as
\begin{align}
    \delta_0 \equiv \text{min}_{n,j}\delta_{j,n}.
\end{align}
is such that in \emph{any} state we obtain $\delta<0$. It is easy to see that for this not to happen we must have:
\begin{align}\label{eq-13fg25}
\delta_0\leq 2(d-1).
\end{align}
In fact, Eq.~\eqref{eq-13fg25} is equivalent to stating that:
\begin{align}\label{eq-41v258}
    \lim_{|\tilde{y}|\to 0} \mathcal{E}(0) \mathcal{E}(\tilde{y}^a) \neq 0,
\end{align}
Note that this is a much weaker constraint than the one discussed in the body of the paper. In fact, it trivially follows from demanding that $\mathcal{E}$ is non-zero as an operator.\footnote{We thank Raghu Mahajan for pointing this out.}.

Let us now investigate how close can $\delta_0$ get to 2(d-1) in a few simple cases. First, in the case of free theories, the leading Regge trajectory ($n=1$) satisfies:
\begin{align}
    \Delta^{\text{even}}_{j=0,1}(J) = d-2+J.
\end{align}
Therefore, from Eq.~\eqref{eq-3jyf2} we have $\delta_{j=0,n=1}=d$, which satisfies the bound \eqref{eq-13fg25} with room to spare in $d\geq3$.

Interestingly, we can obtain $\delta_0 = 2(d-1)$ in planar $\mathcal{N}=4$ SU$(N)$ super Yang-Mills in the large t'Hooft coupling limit.\footnote{We thank David Simmons-Duffin for explaining to us the story of $\mathcal{N}=4$ Super Yang-Mills.} In the $N\to\infty$, limit, we can delineate the Regge trajectories of single and double-trace operators. It turns out that in the large t'Hooft coupling limit, the single-trace $\text{min}_j\Delta^{\text{even}}_{j,1} (J)$ can be arbitrarily increased for each $J$, but the leading double-trace trajectory satisfies $\text{min}_j\Delta^{\text{even}}_{j,1} (J) = 4+J$ (approaching it from below in the $N\to\infty$ limit)~\cite{Hofman:2008ar, Kologlu:2019mfz}. Therefore, the double-trace trajectory leads to $\text{min}_j\delta_{j,1} = 6$, which saturates the bound \eqref{eq-13fg25} given $d=4$.\footnote{For instance, one can consider the trajectory associated with the stress-tensor double-twist $[T T]_J$, which at large spin have twist (i.e., scaling dimension minus spin) $\tau(J) = 2d+J - (J+4) = 2d-4$. We then obtain $\Delta(J=3) = \tau(J=3)+3 = 2d-1$.}

\bibliographystyle{jhep}
\bibliography{bib}

\end{document}